\documentclass[11pt,letterpaper]{article}
\pdfoutput=1
\usepackage{jcappub}
\usepackage{bbm}
\usepackage{mathrsfs}
\usepackage{slashed}
\usepackage{caption}
\usepackage{epstopdf}
\usepackage[normalem]{ulem}
\usepackage[bottom]{footmisc}
\usepackage{subcaption}
\usepackage{bbold}
\usepackage{titlesec}
\usepackage{threeparttable}
\usepackage{booktabs}
\usepackage{changepage}
\usepackage[utf8]{inputenc}
\usepackage{dsfont} 
\usepackage{grffile}
\usepackage{graphicx}  % needed for figures
\usepackage{dcolumn}   % needed for some tables
\usepackage{bm}        % for math
\usepackage{amssymb}   % for math
\usepackage{setspace}
\usepackage{amsmath, amssymb, setspace}
\usepackage{array}
\usepackage{booktabs}
\usepackage{caption}
\usepackage{float}
\usepackage{lmodern}
\usepackage{multirow}
\usepackage{soul}
\usepackage[normalem]{ulem}
\usepackage{braket}
\usepackage{comment}
\usepackage[draft]{pgf}
\usepackage{adjustbox} 
\usepackage{xspace} 
\usepackage{url}
\usepackage{xcolor}
\usepackage{comment}

\usepackage{indentfirst}
\def\aap{Astron.\ Astrophys.\ }

\newcommand{\beq}{\begin{equation}}
\newcommand{\eeq}{\end{equation}}
\newcommand{\bea}{\begin{eqnarray}}
\newcommand{\eea}{\end{eqnarray}}

\newcommand{\gsim}{\lower.7ex\hbox{$\;\stackrel{\textstyle>}{\sim}\;$}}
\newcommand{\lsim}{\lower.7ex\hbox{$\;\stackrel{\textstyle<}{\sim}\;$}}

\newcommand{\be}{\begin{equation}}
\newcommand{\ee}{\end{equation}}
\newcommand{\ba}{\begin{eqnarray}}
\newcommand{\ea}{\end{eqnarray}}

\newcommand{\D}{{\rm d}}

\newcommand{\SNR}{{\rm SNR}}

\newcommand{\hbf}[1]{{\bf \hat{#1}}}
\newcommand{\hbv}[1]{{\bf {#1}}}

\usepackage{soul}

\newcommand{\vev}[1]{\langle #1 \rangle}

\begin{document}

\arxivnumber{DESY-23-201}

\title{\huge{Dissecting the Stochastic Gravitational Wave Background with Astrometry}}

\author{Mesut \c{C}al{\i}\c{s}kan,$^a$}
\author{Yifan Chen,$^b$}
\author{Liang Dai,$^c$}
\author{Neha Anil Kumar,$^a$}
\author{Isak Stomberg$^d$}
\author{and Xiao Xue$^{e,d}$\footnote{Corresponding author.}}
\emailAdd{caliskan@jhu.edu}
\emailAdd{yifan.chen@nbi.ku.dk}
\emailAdd{liangdai@berkeley.edu}
\emailAdd{nanilku1@jhu.edu}
\emailAdd{isak.stomberg@desy.de}
\emailAdd{xiao.xue@desy.de}
\affiliation{$^a$William H. Miller III Department of Physics and Astronomy, Johns Hopkins University, Baltimore, MD 21218, USA\\
$^b$Niels Bohr International Academy, Niels Bohr Institute, Blegdamsvej 17, 2100 Copenhagen, Denmark\\
$^c$Department of Physics, 366 Physics North MC 7300, University of California, Berkeley, CA
94720, USA\\
$^d$Deutsches Elektronen-Synchrotron DESY, Notkestr. 85, 22607, Hamburg, Germany\\
$^e$II. Institute of Theoretical Physics, Universit\"{a}t  Hamburg, 22761, Hamburg, Germany
}

\abstract{Astrometry, the precise measurement of star motions, offers an alternative avenue to investigate low-frequency gravitational waves through the spatial deflection of photons, complementing pulsar timing arrays reliant on timing residuals. Upcoming data from Gaia and Roman can not only cross-check pulsar timing array findings but also explore the uncharted frequency range bridging pulsar timing arrays and LISA. We present an analytical framework to evaluate the feasibility of detecting a gravitational wave background, considering measurement noise and the intrinsic variability of the stochastic background. Furthermore, we highlight astrometry's crucial role in uncovering key properties of the gravitational wave background, such as spectral index and chirality, employing information-matrix analysis. Finally, we simulate the emergence of quadrupolar correlations, commonly referred to as the generalized Hellings-Downs curves.
}

\maketitle

%%%%%%%%%%%%%%%%%%%%%%%
\section{Introduction}
%%%%%%%%%%%%%%%%%%%%%%%

Recent breakthroughs in pulsar timing arrays (PTAs) have heralded a transformative era for the observation of gravitational waves (GWs) at nano-Hertz (nHz) frequencies. These advancements leverage the precise timing information obtained from pulsars, enabling a galactic-scale gravitational wave detector. This innovative approach first led to the detection of a common-spectrum process by the Nanohertz Observatory for Gravitational Waves (NANOGrav) collaboration~\cite{Arzoumanian:2020vkk}, a finding subsequently corroborated by the Parkes PTA, European PTA, and International PTA~\cite{Goncharov:2021oub,Chen:2021rqp,Antoniadis:2022pcn}. More recently, multiple PTA collaborations have found evidence for the quadrupolar angular correlation of these signals on the sky~\cite{NANOGrav:2023gor,EPTA:2023fyk,Reardon:2023gzh,Xu:2023wog}. These collective findings consistently support the presence of a quadrupolar correlation function known as the Hellings-Downs curve~\cite{Hellings:1983fr}, a pivotal characteristic of gravitational-wave induced timing residuals. Furthermore, the amplitude and spectral index of the inferred stochastic gravitational wave background (SGWB) align broadly with predictions derived for a cosmological population of supermassive black hole binaries (SMBHBs) as the gravitational wave sources~\cite{NANOGrav:2023hfp,Antoniadis:2023xlr}. Nonetheless, while not definitively ruled out, certain cosmological sources may still provide potential explanations for the observed SGWB.

On a parallel front, it is essential to note that GWs not only perturb photon travel times but also deflect their paths. This phenomenon serves as the basis for astrometric detection of GWs, which relies on measuring correlated wobbling movements of stars on the sky~\cite{Braginsky:1989pv,Pyne:1995iy,Kaiser:1996wk,Kopeikin:1998ts,Book:2010pf}. Astrometry represents another promising avenue for the detection and characterization of gravitational waves, providing a complementary perspective to the PTA approach~\cite{Moore:2017ity,Klioner:2017asb,Mihaylov:2018uqm,OBeirne:2018slh,Qin:2018yhy,Bini:2018fnv,Darling:2018hmc,Mihaylov:2019lft,Qin:2020hfy,Garcia-Bellido:2021zgu,Aoyama:2021xhj,Wang:2022sxn,Jaraba:2023djs,Liang:2023pbj}. While astrometry currently has less constraining power than PTAs~\cite{Darling:2018hmc,Aoyama:2021xhj,Jaraba:2023djs}, complete datasets from space-borne astrometry missions like Gaia~\cite{2016A&A...595A...1G}, including full time series, have the potential to provide sensitivity comparable to PTAs. {The next-generation astrometric observations have the potential to exceed the sensitivity of PTAs~\cite{Garcia-Bellido:2021zgu}.}

Compared to PTAs, astrometry boasts several noteworthy complementary advantages. The distinct response functions of PTA and astrometry make them sensitive to incoming GWs from different directions as pulsars or stars are more concentrated toward the direction of the Galactic Center~\cite{Moore:2017ity}. Additionally, the strain sensitivity of astrometry remains nearly constant across the frequency spectrum~\cite{Moore:2017ity}, unlike the linear decrease in PTA sensitivity toward higher frequencies. This opens a new window for GWs, exploring uncharted frequency ranges lying between PTAs and the Laser Interferometer Space Antenna (LISA) band~\cite{LISA:2017pwj}. For instance, the Nancy Grace Roman Space Telescope (Roman), with its significantly higher observing cadence, can indeed extend the frequency range to above $10^{-4}$\,Hz~\cite{Wang:2020pmf,Wang:2022sxn,Haiman:2023drc,Pardo:2023cag}. Moreover, the presence of parity-odd correlations among astrometric observables or PTA-astrometry cross-correlations can reveal the existence of a chiral component of SGWB~\cite{Qin:2018yhy,Golat:2022hjf,Liang:2023pbj}. Hence, it is imperative to explore in detail the prospects of astrometry.

This study is dedicated to forecasting the potential of astrometry in discovering SGWBs and characterizing their properties with future data releases, in conjunction with PTAs or without. We establish a framework for predicting the feasibility of astrometric SGWB detection and the resolution of key SGWB parameters. These parameters, including the normalization of characteristic strain, spectral index, and chirality, are crucial for comprehending the distribution of SMBHBs, their potential interaction with the environment, and the presence of any sub-leading cosmological sources. The vector nature of astrometry observables introduces various options for cross-correlations~\cite{Mihaylov:2018uqm,Qin:2018yhy}, in addition to the redshift-only correlation of PTA. Each of these correlations possesses unique quadrupolar correlation functions~\cite{Mihaylov:2018uqm,Qin:2018yhy} and corresponding variances due to the stochastic nature of GWs. Identifying them will not only provide a cross-check of the PTA Hellings-Downs curve at nHz but may also uncover a GW signal at higher frequencies.

The paper's structure is organized as follows: In Sec.\,\ref{sec:basis}, we review the basics of both PTA and astrometry for the detection of the SGWB. Moving to Sec.\,\ref{sec:SNR}, we analytically calculate the sensitivities for various cross-correlation choices associated with PTA and astrometry and evaluate the resolution of key SGWB properties. 
Section\,\ref{sec:correlation} delves into the simulation of spatial correlations, specifically exploring the generalized Hellings-Downs curve, and offers a theoretical insight into intrinsic variance of SGWBs. Finally, in Sec.\,\ref{sec:disucssion}, we draw our conclusions and discuss our findings.

%%%%%%%%%%%%%%%%%%%%%%%%%%%%%%%%%%%%%%%%%%%%%%%%%
\section{PTA and Astrometric Detection of the Stochastic Gravitational Wave Background}\label{sec:basis}
%%%%%%%%%%%%%%%%%%%%%%%%%%%%%%%%%%%%%%%%%%%%%%%%%

%%%%%%%%%%%%%%%%%%%%%%%%%%%%%%%%%%%%%%%%%%%%%%%%%
\subsection{Responses and Angular Correlations for PTA and Astrometry}\label{sec:ang}
%%%%%%%%%%%%%%%%%%%%%%%%%%%%%%%%%%%%%%%%%%%%%%%%%

GWs induce perturbations in the paths of photons along geodesics. These perturbations give rise to two distinct categories of observable phenomena: shifts in the temporal arrival of photons and the proper motion of their sources across the celestial sphere~\cite{Book:2010pf}. PTAs, functioning as an exceptional network of cosmic clocks, possess the remarkable capability to precisely measure the arrival times of radio pulses from distant pulsars. On the other hand, astrometry is dedicated to the precise determination of the positions and motion of stars across the celestial sphere. The shifts depend on both the metric perturbations at the observation point (Earth term) and on the emission sources (pulsar or star terms). However, the latter can usually be disregarded when the distance between the two points significantly exceeds the wavelength of the GWs~\cite{Mihaylov:2018uqm}, or treated as noise when correlations among different baselines result in only the Earth term being coherently summed up. For the purposes of this study, we will focus solely on the Earth term.

The received GW strain at a specific location can be represented as a sum of frequency modes:
\begin{align}
    h_{ij}(t) = \int_{-\infty}^{+\infty} \D f \int \D^2 \hbf{\Omega} \sum_{P}h_P(f,\hbf{\Omega})\,\epsilon_{ij}^P(\hbf{\Omega})\, e^{2\pi i ft}{\,.}\label{Eq:config_def}
\end{align}
In the above equation, $f$, $\hbf{\Omega}$ and $P$ label the frequency, incoming direction, and polarization mode of the GW, respectively. In this work, we will only consider polarization modes within the framework of Einstein's gravity. {The strain amplitude in the frequency domain is given by }$h_P(f,\hbf{\Omega})$, and $\epsilon_{ij}^P(\hbf{\Omega})$ is the polarization basis tensor satisfying $\epsilon_{ij}^P(\hbf{\Omega})  \epsilon^{ij}_{P'}(\hbf{\Omega})=2\delta_{PP'}$. The time-domain strain imposes a real condition, necessitating that $h_P(f,\hbf{\Omega}) = h_P(-f,\hbf{\Omega})^*$ and $ \epsilon_{ij}^{P}(\hbf{\Omega})^*=\epsilon_{ji}^P(\hbf{\Omega})$ for linear polarization basis $P=+/\times$. The frequency-domain signals from PTAs and astrometry can be universally expressed as~\cite{Golat:2022hjf}:
\be X_a(f) \equiv X(f, \hbf{n}_a) = \int_{S^2} \D^2 \hbf{\Omega}\, \sum_P\, h_P(f, \hbf{\Omega})\, \epsilon_{ij}^P(\hbf{\Omega})\,R^{ij}_X(\hbf{\Omega},\hbf{n}_a).\ee
Here, we introduce $X\equiv\{\delta z,\delta \hbv{x} \}$ to encompass both the photon redshift $\delta z$ from PTAs and the proper motion on the celestial sphere $\delta \hbv{x}$ from astrometry. 
The subscript $a$ designates the $a$-th pulsar/star, with the line-of-sight direction denoted as $\hbf{n}_a$. The redshift and astrometric response functions are elucidated in Ref.~\cite{Book:2010pf} as follows:
{\be R_{\delta z}^{ij}(\hbf{\Omega},\hbf{n}) = \frac{1}{2} \left( \frac{\hat{n}^i \hat{n}^j}{ 1+\hbf{\Omega}\cdot\hbf{n}}\right),\qquad R_{\delta {x}_l}^{ij}(\hbf{\Omega},\hbf{n}) = \frac{1}{2} \left[ \frac{\hat{n}^i \hat{n}^j}{ 1+\hbf{\Omega}\cdot\hbf{n}} (\hat{n}^{l} + \hat{\Omega}^{l}) - \delta_{l}^i\hat{n}^j \right].\ee
Here, we utilize a Cartesian coordinate system where the components are labeled by $i,j$ and $l$.}
%labeled as $i$ or $j$ where $l$ represents a direction on the celestial sphere, corresponding to an axis in the Cartesian coordinate system labeled as $i$ or $j$. 
Notably, the timing residual signal represents the time integral of the redshift $\delta z$, introducing an additional factor of $1/(2\pi f)$.

For a Gaussian, stationary GW background, the two-point correlation function of $h_P$ is as follows:
\be
\langle h_P(f,\hbf{\Omega})\,h_{P'}(f', \hbf{\Omega}')^* \rangle = \delta(f-f')\,\delta(\hbf{\Omega},\hbf{\Omega}')\,\mathcal{P}_{PP'}(f,\hbf{\Omega}){\,,} \label{eq:hhC}
\ee
where $\mathcal{P}_{PP'}(f,\hbf{\Omega})$ represents the power spectrum of correlation between the $P$ mode and the $P'$ mode. When the SGWB exhibits isotropy, the power spectrum matrix can be parameterized as~\cite{Sato-Polito:2021efu}:
\begin{align}
 \mathcal{P}_{PP'} (f) =
    \begin{pmatrix}
       I(f)  & -iV(f) \\
        iV(f)& I(f)
    \end{pmatrix}\label{eq:PPP'}
\end{align}
for $P/P' \in \{+,\times\}$, where the real quantities $I(f)$ and $V(f)$ represent the total intensity and the circular polarization, respectively. The isotropic SGWB, as defined in Eqs.\,(\ref{eq:hhC}) and (\ref{eq:PPP'}), results in correlations between two received signals from a pair of pulsars, stars, or pulsar-star pairs, as follows:
\be
\langle X_a(f)\, X_b'(f')^* \rangle = \delta(f-f')
\,\int \D^2 \hbf{\Omega}\,\sum_{P,P'} \,\left(\mathcal{P}_{PP'}(f)\,\epsilon_{ij}^P(\hbf{\Omega})\, \,\epsilon_{kl}^{P'}(\hbf{\Omega})^*\right)\,  R^{ij}_X(\hbf{\Omega},\hbf{n}_a) R^{kl}_{X'}(\hbf{\Omega},\hbf{n}_b){\,.} \label{eq:XXC}
\ee

For PTA with redshift correlations, Eq.\,(\ref{eq:XXC}) leads to the well-known Hellings-Downs curve~\cite{Hellings:1983fr}:
\begin{equation}
    \begin{aligned}
&\langle\delta z_a(f)\delta z_b^*(f')\rangle = \delta(f-f') \, I(f) \, \Gamma_{z}(\theta_{ab}), \\
&\Gamma_{z}(\theta) \equiv \frac{4\pi }{3}\left[
    1-\frac{1}{2}\left(\sin\frac{\theta}{2}\right)^2 + 6\left(\sin\frac{\theta}{2}\right)^2\ln\left(\sin\frac{\theta}{2}\right)
    \right],\label{eq:HD}
        \end{aligned}
\end{equation}
where the function $\Gamma_{z}(\theta_{ab})$ depends solely on $\theta_{ab} \equiv \hbf{n}_a\cdot\hbf{n}_b$ due to the rotational invariance of an isotropic SGWB. 

Correlations involving astrometric motions of a pair of stars or a star and a pulsar can be categorized into directions that are parallel ($\hbf{e}^{a}_{||}$ and $\hbf{e}^{b}_{||}$) and perpendicular ($\hbf{e}_{\perp}$) to their great arc~\cite{Mihaylov:2018uqm}, defined as:
\begin{align}
    \hbf{e}_{\perp} \equiv \frac{\hbf{n}_a\times \hbf{n}_b}{\sqrt{1-(\hbf{n}_a\cdot \hbf{n}_b})^2},\qquad     \hbf{e}^{a}_{||} \equiv
     \frac{\hbf{e}_{\perp}\times\hbf{n}_a}{\sqrt{1-(\hbf{e}_\perp\cdot \hbf{n}_a})^2},
     \qquad \hbf{e}^{b}_{||} \equiv 
     \frac{\hbf{e}_{\perp}\times\hbf{n}_b}{\sqrt{1-(\hbf{e}_\perp\cdot \hbf{n}_b})^2}{\,}. \label{Eq:unit_vectors}
\end{align}
Using these definitions, the correlations in Eq.\,(\ref{eq:XXC}) can be simplified to the following expressions~\cite{Mihaylov:2018uqm,Qin:2018yhy,Golat:2022hjf,Liang:2023pbj}:
\begin{equation}
    \begin{aligned}
        &\langle \delta z_a(f)\delta \hbv{x}_b^*(f') \rangle =\delta(f-f') \, \left(I(f)\,\hbf{e}_{||}^b+iV(f)\,\hbf{e}_{\perp}\right)\,\Gamma_{z\delta \hbv{x}}(\theta_{ab}),\\
        &\langle \delta \hbv{x}_a(f)\delta \hbv{x}_b^*(f')\rangle =\delta(f-f')\,\Big[I(f)\,\left(\hbf{e}_{||}^a\hbf{e}_{||}^b + \hbf{e}_{\perp}\hbf{e}_{\perp}\right)+iV(f)\,\left(\hbf{e}_{||}^a\hbf{e}_{\perp} - \hbf{e}_{\perp}\hbf{e}_{||}^b\right)\Big]\,\Gamma_{\delta \hbv{x}}(\theta_{ab}),
    \end{aligned}
    \label{eq:ORFA}
\end{equation}
where the dimensionless correlation functions satisfy 
\begin{equation}
\begin{aligned}
    \Gamma_{z\delta \hbv{x}}(\theta) \equiv&\,
    \frac{4\pi}{3}\sin(\theta)\left[1+3\left(\tan\frac{\theta}{2}\right)^2\ln\left(\sin\frac{\theta}{2}\right)\right],\\
    \Gamma_{\delta \hbv{x}}(\theta) \equiv&\, \frac{2\pi}{3}
    \left[1 - 7\left(\sin\frac{\theta}{2}\right)^2-12  \left(\sin \frac{\theta }{2}\right)^2\left(\tan \frac{\theta }{2}\right)^2 \ln \left(\sin \frac{\theta }{2}\right)\right].
\end{aligned}\label{Eq:correlations}
\end{equation}
These functions are commonly known as generalized Hellings-Downs curves.

%%%%%%%%%%%%%%%%%%%%%%%%%%%%%%%%%%%%%%%%%%%%%%%%%
\subsection{Gravitational Wave Signal in Spherical Harmonic Space}\label{sec:SH}
%%%%%%%%%%%%%%%%%%%%%%%%%%%%%%%%%%%%%%%%%%%%%%%%%
An alternative representation of SGWB signals in PTAs and astrometric observation is achieved through the use of spherical harmonic space, as demonstrated in prior works such as Refs.~\cite{Gair:2014rwa,Roebber:2016jzl,Qin:2018yhy,Wang:2022sxn,Nay:2023pwu}. 
The key advantage of this representation lies in its ability to diagonalize both the signals and the SGWB-induced variances, simplifying the definition of estimators. 
In this formalism, both the GW-induced redshift and angular deflection can be expressed as discrete summations over the harmonic basis:
\begin{equation}\begin{split}
    \delta z_a(f) =& \sum_{\ell = 2}^{\infty}\sum_{m=-\ell}^{\ell} z_{\ell m}(f)Y_{\ell m}(\hbf{n}_a)\\
   \delta \hbv{x}_a(f) =& \sum_{\ell = 2}^{\infty}\sum_{m=-\ell}^{\ell} \left[E_{\ell m}(f) \hbv{Y}_{\ell m}^E(\hbf{n}_a)+B_{\ell m}(f) \hbv{Y}_{\ell m}^B(\hbf{n}_a)\right]
    \label{eq: z_harm}
    \end{split}
\end{equation}
where $Y_{\ell m}$ represents spherical harmonic functions, and $\hbv{Y}_{\ell m}^E$ and $\hbv{Y}_{\ell m}^B$ are the $E$- and $B$- components of vector spherical harmonic functions, respectively. Correspondingly, $z_{\ell m}(f)$, $E_{\ell m}(f)$, and $B_{\ell m}(f)$ are their respective expansion coefficients. Employing the orthogonality of the spherical harmonic basis, we can determine these components as follows:
\begin{equation}
\begin{split}
    {z}_{\ell m}(f)=& \int_{S^2} \D^2 \hbf{n}\,  \delta z_a(f)\, Y_{\ell m}\left(\hbf{n}\right)^*\,,\\
E_{\ell m}(f) =& \int_{S^2}  \D^2 \hbf{n}\, \delta \hbv{x}_a(f) \cdot \hbv{Y}_{\ell m}^{E}(\hbf{n})^*\,,\\
B_{\ell m}(f) =& \int_{S^2}  \D^2 \hbf{n}\, \delta \hbv{x}_a(f) \cdot \hbv{Y}_{\ell m}^{B}(\hbf{n})^*\,.
\label{eq:zEBr}    \end{split}
\end{equation}
It is important to note that in realistic observations, a finite number of pulsars and non-uniform distributions of pulsars/stars should be taken into account when reconstructing these components. These factors can lead to a mixture of modes and consequently introduce noise~\cite{Roebber:2019gha}. However, recent simulations using mock data have shown that the influence of this mixture is negligible~\cite{Nay:2023pwu}.

To simplify notation, we introduce  $X_{\ell m} \equiv \{z_{\ell m},E_{\ell m},B_{\ell m}\}$ to represent all spherical-harmonic components. With this representation, we can construct the rotationally invariant power spectra:
\be \hbv{C}_{\ell}^{X X'}(f)\equiv\frac{1}{T\,(2 \ell+1)} \sum_{m=-\ell}^{\ell}  X_{\ell m}(f)\, X_{\ell m}'(f)^*\,. \label{Eq:def_cxx}\ee
These power spectra satisfy $\hbv{C}_{\ell}^{XX'}(f)^* = \hbv{C}_{\ell}^{X'X}(f)$. Here, $T$ represents the total observation time, which is used to account for the factor $\delta(0)\rightarrow T$ in the discrete frequency domain.

A convenient way to calculate $X_{\ell m}$ is by using the total angular momentum (TAM) decomposition of GW strain~\cite{Dai:2012bc}, which includes both transverse polarization modes, denoted as $\alpha\in\{TE,TB\}$. Consequently the GW-induced ($h$) spherical harmonic components can be expressed as~\cite{Qin:2018yhy}:
\be
    X_{\ell m}^{(h)}(f) = \sum_\alpha 4\pi i^{\ell}\,F_{\ell}^{X,\alpha}\,h_{\ell m}^{\alpha}(f),\label{eq:Xhlm}
\ee
for $\ell \geq 2$. Here, $F^{X,\alpha}_{\ell}$ are projection factors derived for various combinations of $X$ and $\alpha$~\cite{Qin:2018yhy}. The respective strain amplitudes, denoted as $h_{\ell m}^{\alpha}(f)$, exhibit correlations that can be parameterized as:
\be
\langle h_{\ell m}^{\alpha}(f)\,h_{\ell'm'}^{\beta}(f')^*\rangle = 2\, \delta(f-f')\,
\delta_{\ell\ell'}\,\delta_{mm'}\,\mathcal{P}_{\alpha\beta}(f){\,}.\label{eq:hhTAM}\ee
This assumes an isotropic SGWB with vanishing linear polarization components. The power spectra matrix $\mathcal{P}_{\alpha\beta}$ is structured as~\cite{Kamionkowski:2014faa,Sato-Polito:2021efu}:
\begin{align}
\mathcal{P}_{\alpha\beta}(f) = 
    \begin{pmatrix}
       I(f)  & -iV(f) \\
        iV(f)& I(f)
    \end{pmatrix},
\end{align}
for $\alpha{/\beta} \in \{TE,TB\}$. 

The ensemble average of $\hbv{C}_{\ell}^{X X'}(f)$ comprises an independent sum of GW-induced signals ($h$) and measurement noise ($n$):
\begin{align}
    \langle \hbv{C}_{\ell}^{XX'}(f) \rangle= {C}_{(h)\ell}^{XX'}(f)  +  {C}_{(n)\ell}^{XX'}(f){\,},
    \end{align}
The signal part is directly derived from Eq.\,(\ref{eq:Xhlm}) and (\ref{eq:hhTAM}):
\be
C_{(h)\ell}^{X X'}(f)  = 32\pi^2 \sum_{\alpha,\beta} F_{\ell}^{X,\alpha}\,  F_{\ell}^{X',\beta\,*}\,\mathcal{P}_{\alpha\beta}(f){\,}.\label{Eq:avg_cxx}
\ee
More explicitly, each component of Eq.\,(\ref{Eq:avg_cxx}) is directly related to either the total intensity $I(f)$:
\begin{equation}
\begin{aligned}
        C_{(h)\ell}^{zz}(f) &= I(f)\,A_{\ell},\\
    C_{(h)\ell}^{EE}(f)  =C_{(h)\ell}^{BB}(f) &= I(f)\,A_{\ell}\,B_{\ell}^2  ,\\
    C_{(h)\ell}^{zE}(f) &= I(f)\,A_{\ell}\, B_{\ell},
    \end{aligned}\label{Eq:ChXX_I}
\end{equation}
or the circular polarization $V(f)$~\cite{Qin:2020hfy,Golat:2022hjf,Liang:2023pbj}:
\begin{equation}
\begin{aligned}
    C_{(h)\ell}^{zB}(f)  &= 
    -iV(f)\,A_{\ell}\, B_{\ell},\\
        C_{(h)\ell}^{EB}(f) &= -iV(f)\,A_{\ell}\, B_{\ell}^2.
\end{aligned}\label{Eq:ChXX_V}
\end{equation}
Here, we define
\begin{align}
    A_{\ell} \equiv \frac{16\pi^2}{(\ell+2)(\ell+1)\ell(\ell-1)},\qquad B_{\ell}\equiv \frac{2}{\sqrt{(\ell+1) \ell}}{\,}.
\end{align}

On the contrary, the Gaussian noise inherent in each measurement results in an $\ell$-independent noise spectrum~\cite{Roebber:2016jzl}. In the case of a uniform distribution of $N_X$ pulsars or patches ($N_E = N_B \equiv N_{\delta \mathbf{x}}$) on the celestial sphere, the noise spectra in harmonic space can be expressed as:
\begin{align}
    &\langle X^{(n)}_{\ell m}(f) \, X'^{(n)}_{\ell' m'}(f')^*\rangle = \frac{4 \pi S_X^{(n)}(f)}{ N_X}\,\delta(f-f')\,\delta_{\ell \ell'}\,\delta_{mm'}\,\delta_{XX'}.\label{eq:XXn}
\end{align}
Here, $S_{X}^{(n)}(f)$ represents the noise spectra from each pulsar/patch. For astrometry, these spectra satisfy $S_E^{(n)}(f) = S_B^{(n)}(f) \equiv S_{\delta \mathbf{x}}^{(n)}(f)$, where $S_{\delta \mathbf{x}}^{(n)}(f)$ is the measurement noise of the average proper motion of the patch. The rotational invariant noise spectra directly follow from Eq.\,(\ref{eq:XXn}):
\begin{align}
C_{(n)\ell}^{XX'}(f) = \frac{4\pi S^{(n)}_X(f)}{N_X}\,\delta_{XX'}.\label{Eq:CnXX}
\end{align}

%%%%%%%%%%%%%%%%%%%%%%%%%%%%%%%%%%%%%%%%%%%%%%%%%
\section{Dissecting the Stochastic Gravitational Wave Background}\label{sec:SNR}
%%%%%%%%%%%%%%%%%%%%%%%%%%%%%%%%%%%%%%%%%%%%%%%%%
In this section, we assess the sensitivity of both PTA and astrometry utilizing rotational invariant power spectra $\hbv{C}_{\ell}^{XX'}$. As highlighted earlier, these observables derive considerable advantage from their diagonal nature in harmonic space. Our exploration commences with the formulation of estimators built upon $\hbv{C}_{\ell}^{XX'}$, followed by a thorough analysis of their properties. Subsequently, we employ the information matrix~\cite{doi:10.1098/rsta.1922.0009,Vallisneri:2007ev,Cutler:2007mi} to gauge the sensitivity towards total intensity, spectral index, and chirality.

\subsection{Rotational Invariant Estimators}\label{sec:RIE}
In the frequency domain, discrete frequencies $f_k$ span from $1/T$ to $1/(2\Delta t)$, where $\Delta t$ is the cadence of observation. We use the integer $k$ to label {measurements in a certain frequency bin such that} $\hbv{C}^{XX'}_{\ell,k}\equiv\hbv{C}^{XX'}_{\ell}(f_k)$. The total intensity $I_k$ can be estimated from $XX'=zz/EE/BB/zE$, while the circular polarization $V_k$ arises from the parity-odd observables $zB/EB$. Each estimator is directly constructed from Eq.\,(\ref{Eq:ChXX_I}) and (\ref{Eq:ChXX_V}):
\be\begin{aligned}
\hat{I}_{\ell,k}^{zz} \equiv  \frac{ \hbv{C}_{\ell,k}^{zz} - 4\pi S_{z,k}^{(n)}/N_z}{A_{\ell} }, & \quad \hat{I}_{\ell,k}^{EE/BB} \equiv \frac{\hbv{C}_{\ell,k}^{EE/BB} - 4\pi S_{\delta \hbv{x},k}^{(n)}/N_{\delta \hbv{x}}}{A_{\ell}B_{\ell}^2 },\quad \hat{I}_{\ell,k}^{zE} \equiv 
    \frac{ {\Re}\left[\hbv{C}_{\ell,k}^{zE}\right] }{A_{\ell}B_{\ell} },\\
\hat{V}_{\ell,k}^{EB} & \equiv -\frac{{\Im}\left[\hbv{C}_{\ell,k}^{EB}\right] }{A_{\ell}B_{\ell}^2 },\qquad \hat{V}_{\ell,k}^{zB} \equiv -\frac{{\Im}\left[\hbv{C}_{\ell,k}^{zB}\right] }{A_{\ell}B_{\ell} },    \label{eq:est}
    \end{aligned}
\ee
 whose ensemble averages are either $I_k$ or $V_k$. Given that each power spectrum in harmonic space is independent, the signal-to-noise ratio (SNR) of these estimators in a given frequency bin receives contributions from all achievable $\ell$-modes:
\be
    \begin{aligned}
    \left(\text{SNR}^{XX'}_k\right)^2 =&  \sum_{\ell=2}^{\ell_{\rm max}}
        \frac{2(2\ell+1) \left|C_{(h)\ell,k}^{XX'}\right|^2}{
\left|C_{(h)\ell,k}^{XX'} + \frac{4\pi S_{X,k}^{(n)}}{N_{X}}\delta_{XX'}\right|^2+
\left(C_{(h)\ell,k}^{XX} + \frac{4\pi S_{X,k}^{(n)}}{N_{X}}\right) \left(C_{(h)\ell,k}^{X'X'} + \frac{4\pi S_{X',k}^{(n)}}{N_{X'}}\right)}
        \\
    =&   \sum_{\ell=2}^{\ell_{\rm max}} 
\left\{
\begin{array}{ll}
(2\ell+1)\left(C_{(h)\ell,k}^{XX}/
(C_{(h)\ell,k}^{XX} + 4\pi S_{X,k}^{(n)}/N_{X})\right)^2, & X=X'.\\
  \frac{2(2\ell+1) \left|C_{(h)\ell,k}^{XX'}\right|^2}{
\left|C_{(h)\ell,k}^{XX'}\right|^2+
(C_{(h)\ell,k}^{XX} + {4\pi S_{X,k}^{(n)}}/{N_{X}}) (C_{(h)\ell,k}^{X'X'} + {4\pi S_{X',k}^{(n)}}/{N_{X'}})}, & X\neq X'.
\end{array}
\right.
\end{aligned}
\label{eq:snr}
\ee
Here, $\ell_{\rm max} \sim \sqrt{N_X/2}$ is the highest observable $\ell$ for the constructed power spectra. The denominator corresponds to the variance of the estimator in Eq.\,(\ref{eq:est}), derived using Isserlis' theorem~\cite{10.1093/biomet/11.3.185} for Gaussian fields $X_{\ell m}$.

\begin{figure}[htb]
    \centering
    \includegraphics[width=0.77\textwidth]{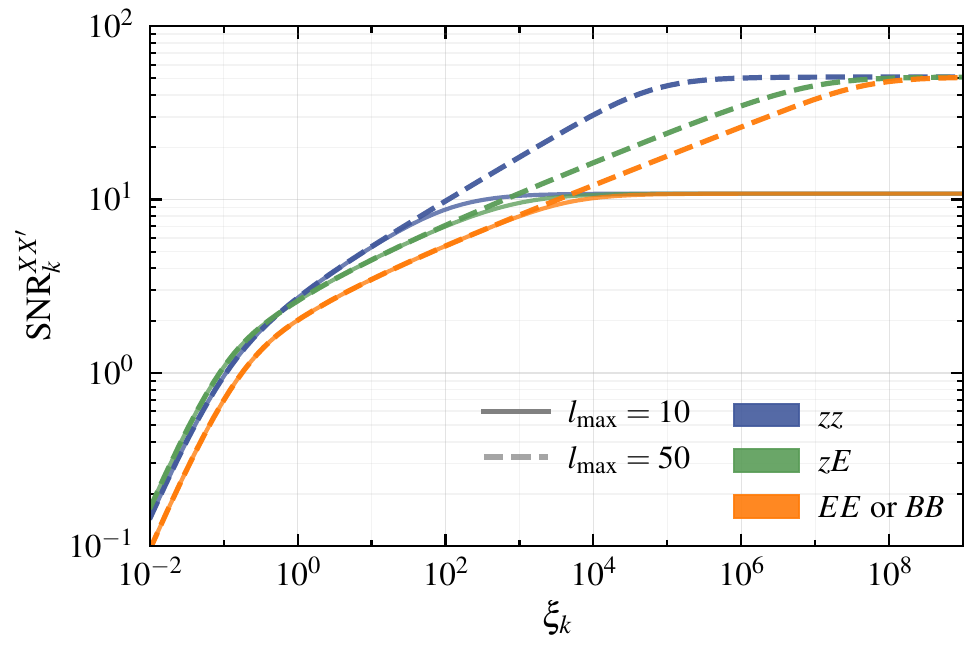} 
    \caption{SNR$^{XX'}_k$ plotted against $\xi_k \equiv {I_k}/{(4\pi S^{(n)}_{X,k}/N_{X})}$ for $I_k$ estimators, encompassing PTA-only correlation $zz$ (blue), astrometry-only correlation $EE$ or $BB$ (orange), and  PTA-astrometry cross-correlation $zE$ (green).
    Two options for the highest $\ell$-mode constructed, $\ell_{\rm max} = 10$ (solid lines) and $50$ (dashed lines), are considered. 
    In the case of $zE$ correlation, we assume $S^{(n)}_{z,k}/N_{z} = S^{(n)}_{E,k}/N_{E}$.
    }
    \label{fig:estGW}
\end{figure}

To gain a quantitative understanding of Eq.\,(\ref{eq:snr}), we depict the SNR$^2_k$ as a function of $\xi_k \equiv {I_k}/{(4\pi S^{(n)}_{X,k}/N_{X})}$ in Fig.\,\ref{fig:estGW} for the $I_k$ estimators. These estimators include PTA-only ($zz$) in blue, astrometry-only ($EE$ or $BB$) in orange, and PTA-astrometry cross-correlation ($zE$) in green. Two scenarios are presented with $\ell_{\rm max} = 10$ (solid lines) and $50$ (dashed lines). The PTA-only case ($zz$) aligns with results previously obtained using cross-correlations in separation-angle space~\cite{Romano:2020sxq}. To better understand the behaviour of the results plotted in Fig.\,\ref{fig:estGW}, we consider the plot categorized into three regions:
\begin{itemize}
\item[$\bullet$] Weak signal ($\xi_k \ll 1$):
The denominator of Eq.\,(\ref{eq:snr}) is dominated by the measurement noise $\sim S_{X,k}^{(n)}$, resulting in SNR$^2 \propto \xi_k^2$.
\item[$\bullet$] Strong signal ($\xi_k \rightarrow \infty$):\\
The measurement noise is subdominant, and the data variance is predominantly determined by the stochastic variance of GW. The saturated value of SNR$^2_k$ is $\sum_{\ell=2}^{\ell_{\rm max}} (2\ell+1) = (\ell_{\rm max}^2 + 2\ell_{\rm max} -3)/2$, which is universal for all cases.
\item[$\bullet$] Intermediate signal:\\
Between the weak and strong signal limits, the contribution of the lowest few $\ell$-modes of $(C_{(h)\ell,k}^{XX'})^2$ surpasses that of the measurement noise. As $C_{(h)\ell,k}^{XX'}$ decreases with higher $\ell$, the transition commences at lower values of $\ell\geq 2$ until the $\ell_\text{max}$-mode becomes dominant. In this region, we introduce $\ell_{\rm eff} \leq \ell_{\rm max}$ to label the highest $\ell$-mode whose power spectra dominates over the measurement noise. Consequently, the SNR$^2_k$ becomes $\sum_{\ell=2}^{\ell_{\rm eff}} (2\ell+1) = (\ell_{\rm eff}^2 + 2\ell_{\rm eff} -3)/2$. $\ell_{\rm eff}$ can be related to $\xi_k$ since $C_{(h)\ell,k}^{XX'} \propto 1/\xi_k$ for large values of $\ell_{\rm eff}$. Thus, for PTA-only with $C_{(h)\ell,k}^{zz} \propto 1/\ell^4$ and astrometry-only $C_{(h)\ell,k}^{EE} = C_{(h)\ell,k}^{BB} \propto 1/\ell^6$, SNR$^2_k$ scales as $\xi_k^{1/2}$ and $\xi_k^{1/3}$, respectively. The $zE$ correlation behaves similarly to astrometry only, as for large $\ell_{\rm eff}$, $C_{(h)\ell,k}^{EE/BB}$ becomes comparable to the measurement noise much later than $C_{(h)\ell,k}^{zz}$.
\end{itemize}
The two cases with different $\ell_{\rm max}$ exhibit nearly identical behavior in the weak signal regions, start to deviate in the middle of the intermediate signal regions, and eventually reach saturation at their respective maximum SNR values.

For estimators of circular polarization $V_k$, the variance incorporates $C_{(h)\ell,k}^{XX}$, which is proportional to $I_k$. Consequently, their maximum SNR is typically suppressed by the ratio of circular polarization $V_k/I_k$.

\subsection{Sensitivity and Parameter Estimation}
In addition to exploring the total intensity and circular polarization, many aspects of the SGWB remain to be unraveled. In this section, we establish an analytical framework to dissect the SGWB. This includes sensitivity estimation and the assessment of various pertinent parameters, all based on the information matrix~\cite{doi:10.1098/rsta.1922.0009}.

We introduce a vector $\hbv{X}_{\ell m,k}$ defined as follows:
\begin{align}
    \hbv{X}_{\ell m,k} \equiv \{z_{\ell m,k},\,E_{\ell m,k},\,B_{\ell m,k}\}^T,
\end{align}
encompassing both redshift and astrometric observables. Assuming Gaussianity and stationarity for both the SGWB and the measurement noise, each
$\hbv{X}_{\ell m,k}$ follows an independent complex multivariate normal distribution:
\begin{align}
     \hbv{X}_{\ell m,k}  \sim \mathcal{CN}\left(\hbv{0}, \hbv{\Sigma}_{\ell,k}\right).\label{Eq:complex_normal}
\end{align}
Here, $\hbv{\Sigma}_{\ell,k}$ is the $3\times 3$ covariance matrix:
\begin{align}
    \renewcommand{\arraystretch}{1.3} % Adjust the number to increase/decrease the spacing
    \hbv{\Sigma}_{\ell,k} =\begin{pmatrix}
        C_{(h)\ell,k}^{zz} & C_{(h)\ell,k}^{zE} & C_{(h)\ell,k}^{zB} \\
        C_{(h)\ell,k}^{zE*} & C_{(h)\ell,k}^{EE} & C_{(h)\ell,k}^{EB} \\
        C_{(h)\ell,k}^{zB*} & C_{(h)\ell,k}^{EB*} & C_{(h)\ell,k}^{BB}
    \end{pmatrix} 
    + \begin{pmatrix}
        C_{(n)\ell,k}^{zz} & & \\
        & C_{(n)\ell,k}^{EE} & \\
        & & C_{(n)\ell,k}^{BB}
    \end{pmatrix},
    \label{eq:CN}
\end{align}
where all the components are defined across Eqs.\,(\ref{Eq:ChXX_I}), (\ref{Eq:ChXX_V}) and (\ref{Eq:CnXX}). 

Given the statistical independence of $\hbv{X}_{\ell m,k}$ for different $m$, $\ell$, and $k$, the joint probability can be expressed as the product of individual probabilities. Consequently, the ln-likelihood function is given by:
\begin{align}
\ln\mathcal{L} (\hbv{X}|\hbv{\mathcal{O}}) = -\sum_{k}\sum_{\ell=2}^{\ell_{\rm max}}\sum_{m=-\ell}^{\ell}\left[
    \hbv{X}_{\ell m,k}^{\dagger}\widetilde{\hbv{\Sigma}}_{\ell,k}(\hbv{\mathcal{O}})^{-1}\hbv{X}_{\ell m,k} + \ln \det{\widetilde{\hbv{\Sigma}}_{\ell,k}(\hbv{\mathcal{O}})}
    \right] + {\rm Const}, \label{Eq:lnlike_general}
\end{align}
where $\hbv{\mathcal{O}}$ represents the model parameters, and $\widetilde{\hbv{\Sigma}}_{\ell,k}(\hbv{\mathcal{O}})$ is the covariance matrix constructed from the corresponding model parameters. Importantly, $\widetilde{\hbv{\Sigma}}_{\ell,k}(\hbv{\mathcal{O}})$ adheres to the condition $\widetilde{\hbv{\Sigma}}_{\ell,k}(\hbv{\mathcal{O}}_{\rm truth}) = \hbv{\Sigma}_{\ell,k}$, as defined in Eq.\,(\ref{eq:CN}), where $\mathcal{O}_{\rm truth}$ represents the true parameters. The ability to estimate the parameters is gauged by the information matrix~\cite{doi:10.1098/rsta.1922.0009}:
\begin{align}
    \mathcal{I}_{ij} \equiv -\left.\left\langle
    \frac{\partial^2 \ln \mathcal{L}}{\partial \mathcal{O}_i\partial \mathcal{O}_j }\right\rangle\right|_{\hbv{\mathcal{O}} = \hbv{\mathcal{O}}_{\rm truth}}.\label{Eq:Fisher_Information}
\end{align}
Here, $\hbv{\mathcal{O}}_i$ represents the $i$-th parameter, and $\langle\dots\rangle$ denotes the ensemble average over $\hbv{X}_{\ell m,k}$. The inverse of the information matrix $\mathcal{I}$ provides the uncertainties for parameter estimation:
\begin{align}
    \sigma^{2}(\mathcal{O}_i) \equiv (\mathcal{I}^{-1})_{ii}.\label{Eq:uncertainty_general}
\end{align}
In cases where the only model parameter of interest is the GW spectrum intensity $I_k$, the SNR of this amplitude is given by:
\begin{align}
    \SNR^2 \equiv   \sum_kI^2_k\sigma^{-2}(I_k).\label{Eq:SNR_general}
\end{align}
It is crucial to note that the information-matrix analysis is accurate primarily when the SNR is robust~\cite{Vallisneri:2007ev,Cutler:2007mi,Ali-Haimoud:2020ozu}. Therefore, we consistently choose $\SNR\geq 1$ as the threshold to ensure the validity of the subsequent discussion.

\subsubsection{Power-law Stochastic Gravitational Wave Background}
In this study, we examine power-law SGWB models and investigate the possible presence of a circularly polarized component. The power-law model is a prevalent approach to characterize the power spectrum of SGWB. Many SGWB models arising from astrophysical or primordial origins can be effectively approximated by power laws in the nHz frequencies. The intensity spectrum of this model is expressed as follows:
\begin{align}
    I(f) = I_{\rm ref} \left(\frac{f}{f_{\rm ref}} \right)^{2\alpha-1}\label{Eq:powerlaw}\,,
\end{align}
where $f_{\rm ref}$ represents the reference frequency, commonly chosen as $f_{\rm ref}=1/{\rm yr}$ in PTA observations, and $\alpha$ denotes the spectrum index. The dimensionless characteristic strain can then be defined as
\begin{align}
h_c(f)\equiv\sqrt{16\pi fI(f)}\equiv A\left(\frac{f}{f_{\rm ref}}\right)^{\alpha}\,,\label{Eq:hc_def}
\end{align}
with $A\equiv (16\pi I_{\rm ref} f_{\rm ref})^{1/2}$ serving as the strain amplitude normalization factor at $f_{\rm ref}$. This normalization choice is in accordance with Refs.~\cite{Allen:1996vm,EPTA:2015qep}.

The spectral index of the SGWB can deviate from a constant value. At the low-frequency end of the nHz spectrum, potential interactions with the environment~\cite{Armitage:2002uu,Sesana:2004sp,Merritt:2004gc} or orbital eccentricities of SMBHBs~\cite{Enoki:2006kj} lead to a turning of the spectrum slope~\cite{NANOGrav:2023hfp}. The high-frequency turning, occurring around $\sim 10^{-6}$\,Hz for SMBHBs with masses of $\sim 10^9\,M_{\odot}$, happens as SMBHBs approach the merger phase~\cite{Ajith:2007kx}. However, in the frequency range we consider, the low-frequency deviation is expected to be small, while the high-frequency contribution is sub-leading in sensitivity. Thus, we focus on a constant spectral index in the following.

Another feature of the SGWB is chirality, parameterized by macroscopic circular polarization. Cosmological models, such as those described in Refs.~\cite{Lue:1998mq,Garcia-Bellido:2016dkw,Obata:2016oym,Machado:2018nqk,Qiao:2022mln}, can generate chirality through parity-violating interactions. A finite sum of nearby SMBHBs can also produce a random fraction of chirality~\cite{Conneely:2018wis,Hotinli:2019tpc,Ellis:2023owy}. As $C_{(h)\ell,k}^{zz}$ in Eq.\,(\ref{Eq:ChXX_I}) is dependent solely on $I_k$ and lacks any $V_k$ dependence, measuring the isotropic circular polarization map using PTA-only observations is not possible. It is noteworthy that PTA can still probe anisotropic circular polarization~\cite{Kato:2015bye,Belgacem:2020nda,Tasinato:2023zcg,AnilKumar:2023yfw}, which is beyond the scope of this study.

\paragraph{Fiducial SGWB Model from PTA Observation}

Recent observations by NANOGrav~\cite{NANOGrav:2023gor}, PPTA~\cite{Reardon:2023gzh}, EPTA~\cite{EPTA:2023fyk}, and CPTA~\cite{Xu:2023wog} suggest that the SGWB signal is consistent with that produced by SMBHBs. Assuming that the nearly-circular orbit evolution is predominantly driven by GW emission, the SGWB spectrum can be well-described by a power-law model with $\alpha =-2/3$, despite a potential deviation at the low-frequency end due to environmental effects or eccentric orbits~\cite{Chen:2016zyo,NANOGrav:2023hfp,Ellis:2023dgf}. With $\alpha=-2/3$ fixed, the strain found by NANOGrav (NG) is given by
\begin{align}
   h_{c}^{\rm NG}(f) =  A_{\rm NG} \left(\frac{f}{f_{\rm ref}}\right)^{-2/3}\,.\label{Eq:Phi_to_PSD1}
\end{align}
where $A_{\rm NG}\simeq 2.4\times 10^{-15}$~\cite{NANOGrav:2023gor}.  Utilizing Eq.\,(\ref{Eq:hc_def}), the power spectrum of our fiducial SGWB model has a reference intensity $I^{{\rm NG}}_{\rm ref}=(1/16\pi)A_{\rm NG}^2\,{\rm yr} \simeq 1.1\times 10^{-31}$\,yr. In the subsequent sections, the fiducial model will be employed to compare sensitivities across different observational channels. Any deviation from $\alpha = -2/3$ can be interpreted as environmental effects or primordial origins.

\paragraph{Power-law Model Parameters}
The power-law model involves two parameters, denoted as $\mathcal{O}=\{\log_{10}A,\alpha\}$. The derivatives with respect to these parameters are obtained through the chain rule:
\begin{equation}
\begin{aligned}
    &\frac{\partial}{\partial \log_{10}A} =2\ln(10)\sum_k
    I_k
    \frac{\partial}{\partial I_k},\\
    &\frac{\partial}{\partial \alpha} =\sum_k 2\ln\left(k\Delta_f\right)I_k\frac{\partial}{\partial I_k},
\end{aligned}
\end{equation}
where we define the dimensionless factor $\Delta_f\equiv 1/(T f_{\rm ref})$. The  information matrix is expressed as
\begin{align}
    \mathcal{I}= 4\sum_k \frac{I_k^2}{\sigma_k^2}
    \begin{pmatrix}
   \ln(10)^2&\ln(k\Delta_f)\ln(10)\\
   \ln(k\Delta_f)\ln(10) &\ln(k\Delta_f)^2
    \end{pmatrix},
\end{align}
where $\sigma_k\equiv \sigma(I_k)$. Note that the matrix inside the summation may appear singular, but the overall matrix after summation is not. The uncertainties of $\{\log_{10}A,\alpha\}$ are given by
\begin{equation}
\begin{aligned}
&\sigma^{-2}(\log_{10}A) = 4\ln(10)^2\left[\sum_k \frac{I_k^2}{\sigma_k^2} - \frac{\left[\sum_k I_k^2/\sigma_k^2\,\ln(k\Delta_f)\right]^2}{\sum_k I_k^2/\sigma_k^2\,\ln(k\Delta_f)^2}\right],\\
&\sigma^{-2}(\alpha) = 4\left[\sum_k \frac{I_k^2}{\sigma_k^2}\,\ln(k\Delta_f )^2 - \frac{\left[\sum_k I_k^2/\sigma_k^2\, \ln(k\Delta_f )\right]^2}{\sum_k I_k^2/\sigma_k^2}\right],\label{Eq:PL_estimation}
\end{aligned}
\end{equation}
respectively. An alternative parameterization using $\mathcal{O}=\{\log_{10}A,\gamma\}$, where $\alpha \equiv (3-\gamma)/2$, results in a straightforward rescaling of $\sigma(\gamma )= 2\sigma(\alpha)$.

The assessability of both $A$ and $\alpha$ is significantly dependent on the total $\SNR$. However, the weighted summation of individual frequency bins adds complexity to the equations. In the following, we will directly calculate $\sigma^{-2}(\alpha)$ and $\sigma^{-2}(\log_{10}A)$ and examine their behavior in different observation channels.

\subsubsection{Pulsar Timing Arrays}
In PTA-only observations, the ln-likelihood defined in Eq.\,(\ref{Eq:lnlike_general}) exclusively involves the redshift $z_{\ell m,k}$, obtained by marginalizing over all $E_{\ell,m}$ and $B_{\ell,m}$:
\begin{align}
    \ln\mathcal{L}_{\rm (PTA)} = -\sum_{k}\sum_\ell^{\ell_{\rm max}}\sum_{m=-\ell}^{\ell}\left[\frac{z_{\ell m,k}^* z_{\ell m,k}}{C_{(h)\ell,k}^{zz} + C_{(n)\ell,k}^{zz}} + \ln \left(C_{(h)\ell,k}^{zz} + C_{(n)\ell,k}^{zz}\right)\right] +{\rm Const}.
\end{align}
As discussed previously, this likelihood is not sensitive to $V_k$.

According to Eq.\,(\ref{Eq:uncertainty_general}), the uncertainty of each $\{I_k\}$ is given by
\begin{align}
    \sigma^{-2}_k=((\mathcal{I}^{-1})_{kk})^{-1}=
    \sum_{\ell=2}^{\ell_{\rm max}}(2\ell+1)\left[
    \frac{4\pi S^{(n)}_{z,k}}{N_{z} A_{\ell}} + I_k
    \right]^{-2}. \label{Eq:PTA_Ierr}
\end{align}
The total SNR follows Eq.\,(\ref{Eq:SNR_general})
\begin{align}
   \SNR_{\rm (PTA)}^2 =\sum_k \sum_{\ell=2}^{\ell_{\rm max}}(2\ell + 1)\left[
   (\xi_k^z A_{\ell})^{-1} + 1
   \right]^{-2},\qquad
   \xi_k^z \equiv \frac{I_k}{4\pi S^{(n)}_{z,k}/N_{z}},\label{Eq:SNR_PTA}
\end{align}
which aligns with the estimator using $\hbv{C}_{\ell,k}^{zz}$ in Eq.\,(\ref{eq:snr}). Given various $\xi_k^z$ value ranges, the SNR can be categorized into weak, intermediate, and strong signal regions, as discussed in Sec.\,\ref{sec:RIE}.

\paragraph{PTA Sensitivity to the Power-law SGWB} We proceed to estimate the sensitivity to the power-law SGWB in realistic PTA observations. The Gaussian noise for timing residual consists of a white noise component and a red noise one, relatively well-fitted by the power-law spectrum~\cite{NANOGrav:2023ctt}. By transitioning from timing residual (TR) to redshift, the noise spectrum for a single pulsar can be expressed as
\begin{align}
    S_{z}^{(n)}(f) = S_{z}^{(r)}(f_{\rm ref})  \left(\frac{f}{f_{\rm ref}} \right)^{\gamma_r}  + S_{z}^{(w)}(f_{\rm ref})  \left(\frac{f}{f_{\rm ref}} \right)^{2},\quad S_{z}^{(w)}(f_{\rm ref}) \equiv \sigma_{\rm TR}^2 (2\pi f_{\rm ref})^2 \Delta t. \label{Eq:PTA_RN}
\end{align}
Here, $S_{z}^{(r)}$ and $S_{z}^{(w)}$ are the corresponding noise components, $\sigma_{\rm TR}$ is the timing residual uncertainty of each measurement for a pulsar, $\Delta t$ is the cadence of the observation.
%, and $T_{\rm obs}$ is the total observation time.

We consider recent NANOGrav and future SKA observations. The red noise for NANOGrav is fit to be $S_{z}^{(r)}  = 1.3\times 10^{-28}$\,yr and $\gamma_r = 0$~\cite{NANOGrav:2023ctt,NANOGrav:2023gor}, while for SKA, we assume there is only white noise. The benchmark parameters for the two observations are taken from~\cite{NANOGrav:2023ctt} and \cite{Moore:2014eua,Moore:2014lga}, respectively:
\be
\begin{aligned}
\text{NANOGrav}:&\qquad \sigma_{\rm TR}=80\, \text{ns},\ \Delta t = 14\, {\rm days},\ T_{\rm obs} = 15\,{\rm yr},\ N_{z} = 50,\\
\text{SKA}:&\qquad \sigma_{\rm TR}=30\,\text{ns},\ \Delta t = 14\, {\rm days},\ T_{\rm obs} = 20\,{\rm yr},\ N_{z} = 200.
\label{Eq:PTA_obs}
\end{aligned}\ee
Here, $T_{\rm obs}$ is the total observation time.
Comparing our noise model for NANOGrav with the fiducial SGWB model defined in Eq.\,(\ref{Eq:powerlaw}), we find that the SGWB dominates the noise in the lowest $5$ bins, consistent with NANOGrav's result~\cite{NANOGrav:2023gor}. With $N_z = 50$ pulsars for NANOGrav, this leads to $\ell_{\rm max} = 5$, while for SKA, $\ell_{\rm max}= 10$.

\begin{figure}[ht]
    \centering
\includegraphics[width=\textwidth]{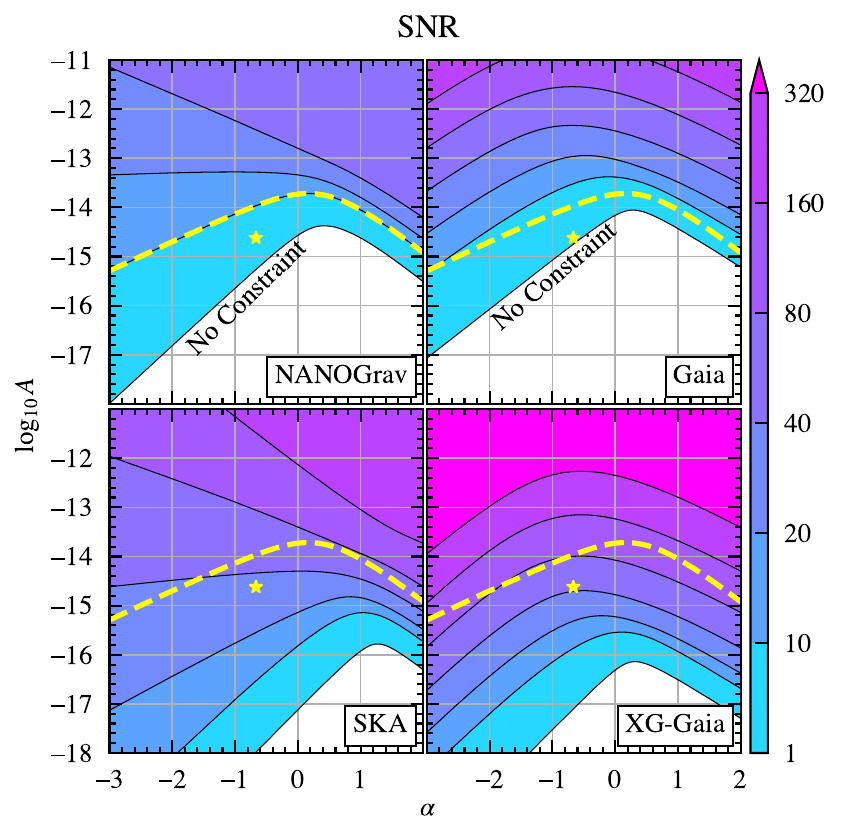}   
    \caption{The SNR distribution as a function of the power-law SGWB model parameters, the reference strain $\log_{10}A$ and spectral index $\alpha$ as defined in Eq.\,(\ref{Eq:hc_def}), for the four considered observations. The left two correspond to PTAs, while the right two represent astrometric observations. The white regions indicate $\text{SNR}<1$, where the  information matrix is not applicable. Yellow stars mark the fiducial parameters from Eq.\,(\ref{Eq:Phi_to_PSD1}), with SNR values of $4.0$, $1.5$, $34.1$, and $43.9$ for NANOGrav, Gaia, SKA, and {XG-Gaia}, respectively. The current exclusion region, defined as $\text{SNR}>10$ for NANOGrav, is highlighted by yellow dashed lines.}    
    \label{fig:SNR_PTA}
\end{figure}

In the left part of Fig.\,\ref{fig:SNR_PTA}, we depict the SNR distribution of the power-law SGWB as a function of $\log_{10}A$ and $\alpha$ for both PTA observations. The fiducial SGWB model yields an SNR\,$\simeq 4.0$ for NANOGrav (NG) and SNR\,$\simeq 34.1$ for SKA. Notably, we observe that above $\SNR\approx 80$, the NANOGrav sensitivity reaches a saturation phase, where a stronger $A$ no longer improves the SNR due to the intrinsic variation in SGWB, consistent with the result in Ref.~\cite{Romano:2020sxq}. On the other hand, for SKA with a larger value of $\ell_{\rm max}$ and more sensitive frequency bins, the threshold for saturation is significantly higher.

\begin{figure*}[ht]
    \centering
\includegraphics[width=1\textwidth]{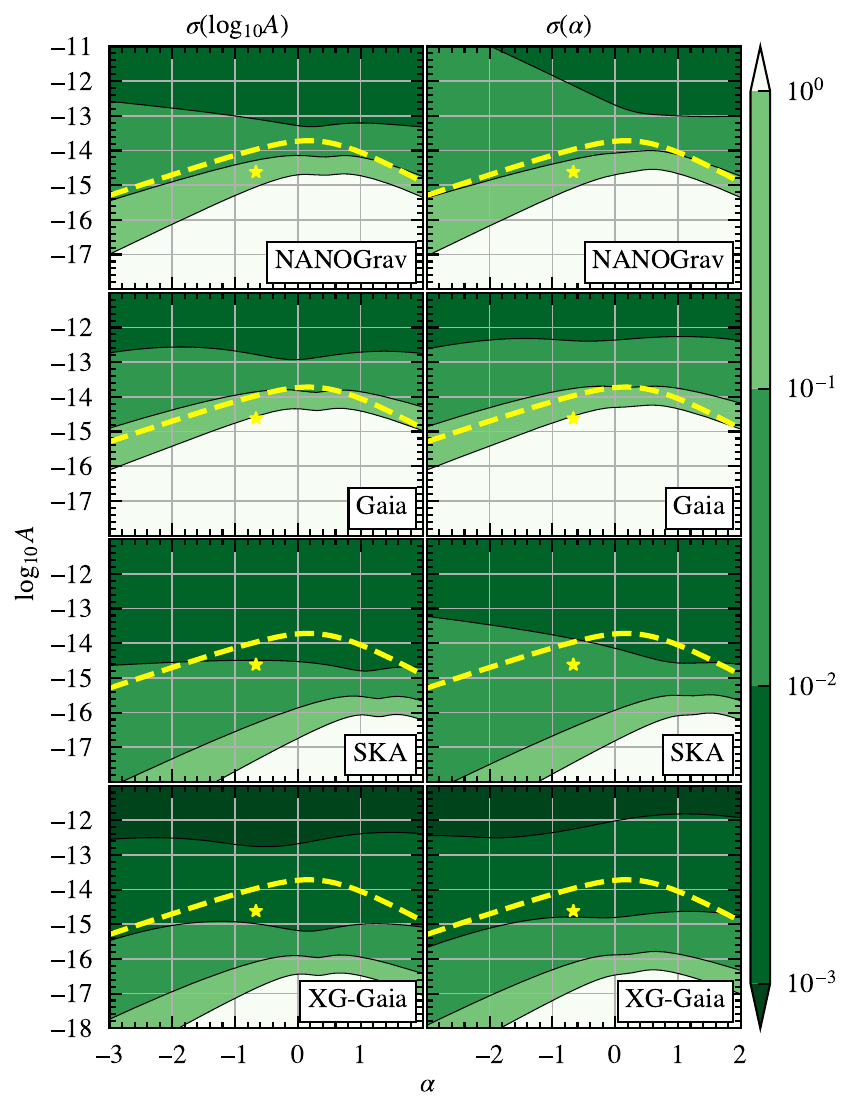}
\caption{The resolution of the two power-law SGWB model parameters, $\sigma(\log_{10}A)$ and $\sigma(\alpha)$, as a function of $\log_{10}A$ and $\alpha$, for the four considered observations. The yellow stars correspond to the fiducial parameters observed by NANOGrav, and the yellow dashed lines are excluded in its $\text{SNR}>10$ region. The resolutions at the fiducial value, from top to bottom, are $\sigma(\log_{10}A) = 0.25$, $1.3$, $0.012$, and $0.0056$, and $\sigma(\alpha) = 0.24$, $1.3$, $0.017$, and $0.0080$.}
    \label{fig:sensitivity_para}
\end{figure*}

We also compute the estimation uncertainty of the power-law model parameters, $\log_{10}A$ and $\alpha$, using Eq.\.(\ref{Eq:uncertainty_general}), as illustrated in Fig.\,\ref{fig:sensitivity_para}. For the fiducial SGWB model, the NANOGrav case yields $\sigma(\log_{10}A)\simeq0.25$ and $\sigma(\alpha)\simeq0.24$, consistent with their data analysis~\cite{NANOGrav:2023gor}. In comparison, the SKA demonstrates superior resolution over NANOGrav by factors of $21$ and $14$ for the two parameters, respectively. Akin to the total SNR observations, the uncertainty ceases to improve beyond the saturation phase for NANOGrav.

\clearpage

\subsubsection{Astrometry}\label{sec:ast}
For astrometry, we have two independent measurements, $\hbv{X}_{\ell m,k}=\{E_{\ell m,k},B_{\ell m,k}\}$, resulting in a covariance matrix:
\begin{align}
    \renewcommand{\arraystretch}{1.3}
    \hbv{\Sigma}_{\ell,k} =
    \begin{pmatrix}
        C_{(h)\ell,k}^{EE}&C_{(h)\ell,k}^{EB}\\
        C_{(h)\ell,k}^{EB*}&C_{(h)\ell,k}^{BB}
    \end{pmatrix} 
    + \begin{pmatrix}
        C_{(n)\ell,k}^{EE}&\\
        &C_{(n)\ell,k}^{BB}
    \end{pmatrix}.\label{Eq:Astro_Cov}
\end{align}
Here, $C_{(h)\ell,k}^{EE/BB}$ is proportional to $I_k$, while $C_{(h)\ell,k}^{EB}$ is proportional to $V_k$. We insert the covariance matrix Eq.\,(\ref{Eq:Astro_Cov}) into the likelihood function Eq.\,(\ref{Eq:lnlike_general}), labeling it as $\ln\mathcal{L}_{\rm (ast)}$. 

 We first calculate the $\SNR$ of the total intensity $I_k$ assuming $v_k \equiv V_k/I_k = 0$. The measurement uncertainty of $I_k$ follows Eq.\,(\ref{Eq:uncertainty_general}),
\begin{align}
     \lim_{v_k\xrightarrow{}0}\sigma^{-2}(I_k)=  2\sum_{\ell=2}^{\ell_{\rm max}}(2\ell+1) \left[
    \frac{4 \pi  S_{\delta \mathbf{x},k}^{(n)}}{N_{\delta \mathbf{x}}A_{\ell}B_{\ell}^2} + I_k 
    \right]^{-2}\label{Eq:Astro_sigma}.
\end{align}
The total SNR is derived from Eq.\,(\ref{Eq:SNR_general}):
\begin{align}
   &\SNR_{\rm (ast)}^2 = 2\sum_k\sum_{\ell=2}^{\ell_{\rm max}}(2\ell + 1)
    \left[
   (\xi_{k}^{\delta \mathbf{x}} A_{\ell}B_{\ell}^2)^{-1} + 1
   \right]^{-2},\qquad \xi_{k}^{\delta \mathbf{x}}\equiv \frac{I_k}{4\pi S^{(n)}_{{\delta \mathbf{x}},k}/N_{\delta \mathbf{x}}}.
   \label{Eq:Gaia_SNR}
\end{align}
This expression is equivalent to the sum of estimators using $\hbv{C}_{\ell,k}^{EE}$ and $\hbv{C}_{\ell,k}^{BB}$ in Eq.\,(\ref{eq:snr}). The SNR is very similar to that of PTA in Eq.\,(\ref{Eq:PTA_Ierr}), despite the additional factor $B_{\ell}^2 \propto 1/(\ell(\ell+1))$, which makes higher $\ell$-modes more suppressed. Consequently, while astrometry can accommodate a significantly higher $\ell_{\rm max}$ due to the larger number of stars compared to pulsars, redistributing stars into an appropriate number of patches will not diminish sensitivity.

\paragraph{Astrometric Sensitivity to the Power-law SGWB}

Due to the large number of observed stars in astrometry, an efficient strategy, without sacrificing sensitivity, involves dividing the celestial sphere into various patches~\cite{Moore:2017ity, Klioner:2017asb}. The distribution of these patches can be realized through \texttt{HEALPix}~\cite{Gorski:2004by}. Each patch combines the spatial deflections of stars on the celestial sphere into an average $\delta \hbv{x}$. For $N_S$ stars evenly distributed in $N_{\delta \hbv{x}}$ patches, the two-sided noise power spectrum of each patch is given by
\begin{align}
    S_{\delta \hbv{x}}^{(n)} = \frac{\sigma_{\theta}^2\Delta t}{N_{S}/N_{\delta \hbv{x}}}.
    \label{eq:Sndeltax}
\end{align}
Here, $\sigma_\theta$ represents the uncertainty of each measurement for a star. The SNR contribution from a specific $\ell$-mode is characterized by $\xi_{k}^{\delta \mathbf{x}}$ in Eq.\,(\ref{Eq:Gaia_SNR}), which is independent of the number of patches $N_{\delta \hbv{x}}$. On the other hand, $N_{\delta \hbv{x}}$ determines $\ell_{\rm max}$, influencing SNR in the strong signal region. Thus, a reasonable choice of $N_{\delta \hbv{x}}$ is necessary in astrometric observation.

Gaia~\cite{2016A&A...595A...1G} and upcoming missions such as Roman~\cite{Wang:2020pmf, Wang:2022sxn, Haiman:2023drc, Pardo:2023cag} and Theia~\cite{Theia:2017xtk,Malbet:2021rgr} play pivotal roles in astrometric observations. The Gaia mission, having measured the proper motion of over $\sim 10^9$ stars and $\sim 10^6$ quasi-stellar objects (QSOs) for more than $10$ years, has provided a rich dataset. The QSO data from Gaia has been leveraged to constrain ultralow ($\ll$\,nHz) frequency gravitational waves~\cite{Darling:2018hmc, Aoyama:2021xhj, Jaraba:2023djs}. In this study, we consider the full dataset from Gaia upon its release, including its comprehensive measurements. {The proposed mission Theia boasts significantly improved resolution but with a small field of view~\cite{Theia:2017xtk,Malbet:2021rgr}. We anticipate that the next-generation upgrade of Gaia, which we abbreviate as 'XG-Gaia', will achieve $\mu$as-resolution while maintaining its cadence and the number of observed stars~\cite{10.3389/fspas.2018.00011}. The benchmark parameters for the two astrometric missions are listed as follows~\cite{Moore:2017ity,10.3389/fspas.2018.00011}:
\begin{equation}
\begin{aligned}
\text{Gaia}:&\qquad    \sigma_{\theta}=100 \,\mu{\rm as},\
\Delta t = 24\,\text{days},\ T_{\rm obs} = 10\,{\rm yr},\ 
N_S=1.5\times 10^9,\\
{\text{XG-Gaia}}:&\qquad   \sigma_{\theta}= 1\,\mu{\rm as},\
\Delta t = 24\,\text{days},\ T_{\rm obs} = 20\,{\rm yr},\ 
N_S=1.5\times 10^{9}. 
\label{Eq:Gaia_obs}
\end{aligned}
\end{equation}
Note that we adopt a conservative estimate for the number of stars compared to Ref.~\cite{Garcia-Bellido:2021zgu}.}

We evaluate the SNR for various measurements of the power-law SGWB model, assuming uniform distribution of stars across the celestial sphere, each possessing identical measurement properties. While a total of $N_S=1.5\times10^9$ stars would allow for an $\ell_{\rm max}\sim 10^4$, we opt for $N_{\delta \hbv{x}}=40000$ for Gaia and {XG-Gaia} with $\ell_{\rm max}=200$, as further increasing the number no longer significantly enhances the SNR within our parameter space of interest.

In the right panel of Fig.\,\ref{fig:SNR_PTA}, we depict the SNR distribution of Gaia and {XG-Gaia} using Eq.\,(\ref{Eq:Gaia_SNR}) concerning power-law model parameters. Generally, PTA's sensitivity exhibits a more pronounced change in terms of $\alpha$, attributed to its more tentative decrease in sensitivity at higher frequencies. The fiducial SGWB signal yields an ${\rm SNR}\simeq 1.5$ for Gaia, compared to $4.0$ in NANOGrav, indicating Gaia resides in the marginal region for cross-checking PTA discoveries. On the other hand, {XG-Gaia} achieves an ${\rm SNR}\simeq 43.9$, higher than SKA, benefiting from contributions at higher $\ell$ and frequency modes.

In Fig.\,\ref{fig:sensitivity_para}, we compare the resolution of power-law parameters in astrometry with those in PTA. The resolution approximately follows the distribution of $1/\text{SNR}$. {XG-Gaia}, with resolutions below $10^{-2}$ for both amplitude and spectral index, can provide an exceptionally precise dissection of the SGWB spectrum. This high precision is pivotal for gaining insights into the distribution and evolution of SMBHBs, including potential environmental effects.

The Roman telescope, distinguished by its much higher cadence, demonstrates sensitivity to the SGWB in the frequency band situated between PTA and LISA~\cite{Wang:2022sxn,Pardo:2023cag}. The associated frequency range of Roman offers valuable insights into whether the SGWB displays a frequency turning point higher than the typical PTA band. This aspect is crucial for elucidating the mass distribution of SMBHBs and investigating potential cosmological components of the SGWB. Additionally, it allows for cross-correlation with other proposed measurements in the same frequency band, such as binary neutron star resonance~\cite{Blas:2021mpc,Blas:2021mqw}.

\paragraph{Identifying Circular Polarization with Astrometry}

Astrometry holds the potential to explore circular polarization through $EB$ correlation~\cite{Qin:2018yhy,Golat:2022hjf,Liang:2023pbj}. We estimate the resolution of the circular polarization fraction $v_k \equiv V_k/I_k$, considering parameters on each frequency mode $\mathcal{O}_k\equiv\{I_k,v_k\}$. The corresponding information matrix, as per Eq.\,(\ref{Eq:Fisher_Information}), is given by
\begin{equation}
\begin{aligned}
&\mathcal{I}_{k}^{ij} 
=\frac{1}{I_k^2}\sum_{\ell=2}^{\ell_{\rm max}}(2\ell+1)
\frac{ 2\zeta_{\ell,k}^2}{\left(1+2 \zeta_{\ell,k} +\zeta_{\ell,k} ^2 u_k \right)^2}\\
    &\qquad\times\begin{pmatrix}
       2-u_k+2\zeta_{\ell,k}  u_k+\zeta_{\ell,k} ^2 u_k
   &I_k v_k \left(1-\zeta_{\ell,k} ^2 u_k\right)\\
        I_kv_k \left(1-\zeta_{\ell,k} ^2 u_k\right)&I_k^2\left(1+2 \zeta_{\ell,k} +\zeta_{\ell,k} ^2 \left(2-u_k\right)\right)
\end{pmatrix},
\label{eq:FIMIv}
\end{aligned}
\end{equation}
where $u_k \equiv 1-v_k^2 $ and $\zeta_{\ell,k} \equiv \xi_{k}^{\delta \mathbf{x}} A_{\ell}B_{\ell}^2 = I_kA_{\ell}B_{\ell}^2/(4\pi S_{\delta \hbv{x},k}^{(n)}/N_{\delta \hbv{x}})$.

Due to the complexity of the expression in Eq.\,(\ref{eq:FIMIv}), we consider a small portion of circular polarization in the limit $|v_k|\ll 1$. In this limit, the off-diagonal terms $\mathcal{I}_k^{vI}$ are sub-leading, implying that $I_k$ and $v_k$ are uncorrelated. We find the inverse of the uncertainties as
\begin{equation}
\begin{aligned}
    &\sigma^{-2}(I_k)  = \left(\mathcal{I}_{k}^{-1}\right)_{II} = \frac{\mathcal{I}_k^{vv}}{\mathcal{I}_k^{vv}\mathcal{I}_k^{II}-(\mathcal{I}_k^{Iv})^2} \qquad \xrightarrow{|v_k|\ll1} \qquad \frac{2}{I_k^2}\sum_{\ell}^{\ell_{\rm max}}(2\ell+1)(1+\zeta^{-1}_{\ell,k})^{-2}, \\
    &\sigma^{-2}(v_k)  = \left(\mathcal{I}_{k}^{-1}\right)_{vv} = \frac{\mathcal{I}_k^{II}}{\mathcal{I}_k^{vv}\mathcal{I}_k^{II}-(\mathcal{I}_k^{Iv})^2} \qquad \xrightarrow{|v_k|\ll1} \qquad 2\sum_{\ell}^{\ell_{\rm max}}(2\ell+1)(1+\zeta^{-1}_{\ell,k})^{-2}.\label{Eq:Astro_Circ}
\end{aligned}
\end{equation}
Here, the uncertainty for $I_k$ aligns with the expression in Eq.\,(\ref{Eq:Astro_sigma}), as expected.

\begin{figure}[t]
    \centering
\includegraphics[width=1.0\textwidth]{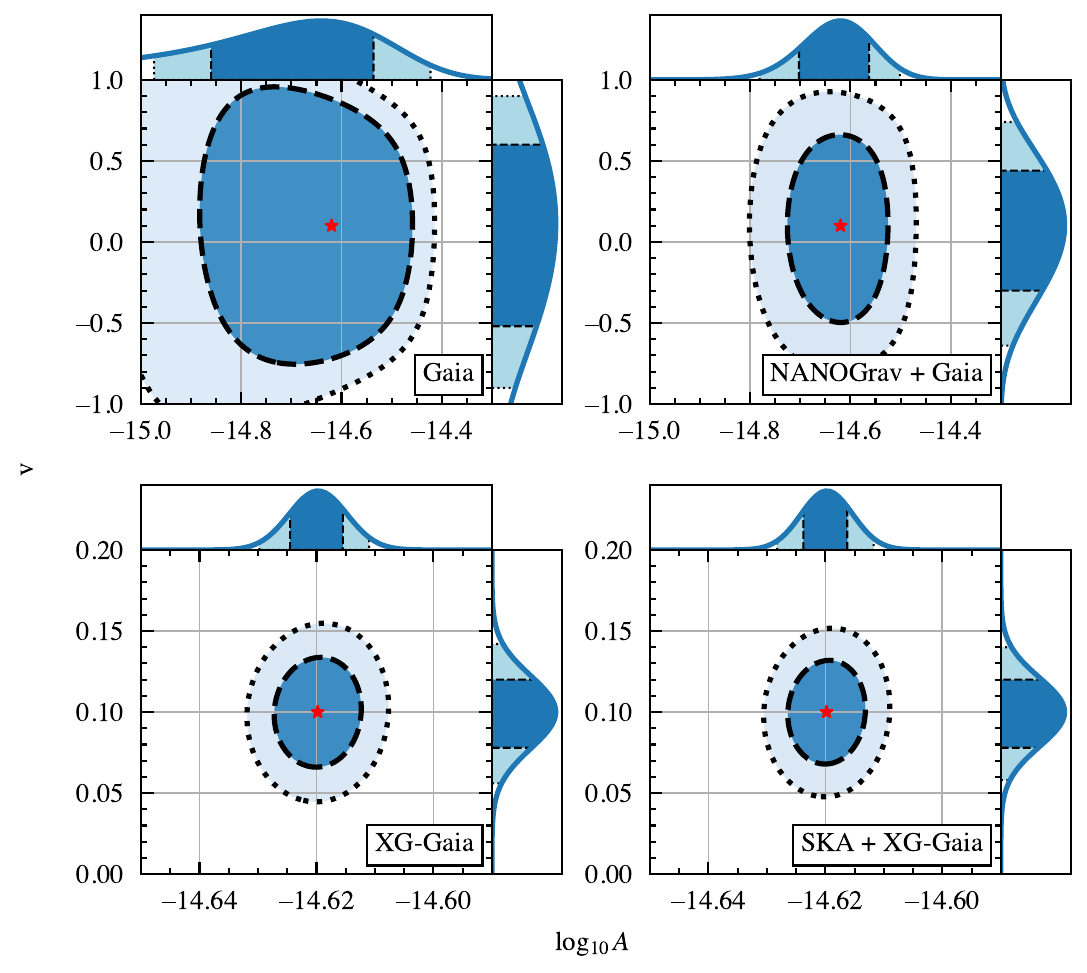} 
    \caption{
    Posteriors of the reference strain amplitude $\log_{10}A$ and the fraction of circular polarization $v\equiv V_k/I_k$, for astrometric observations (left) and PTA-astrometry synergistic analyses (right). The red stars correspond to the true parameters, including the fiducial value for $\log_{10}A$ and the assumed $v=0.1$. The marginalized distribution of each parameter is presented next to the posterior distribution. The dark and light blue regions represent the $1\sigma$ and $2\sigma$ regions, respectively. The $1\sigma$ uncertainties for $v$ are $\sigma(v) = 0.57$, $0.39$, $0.022$, and $0.022$ for Gaia, NANOGrav\,$+$\,Gaia, {XG-Gaia}, and SKA$\,+\,${XG-Gaia}, respectively.}
    \label{fig:Posterior_Astro}
\end{figure}

Considering a power-law model with $\alpha$ fixed at the fiducial value and assuming a constant circular polarization fraction $v_k$ across all frequencies (i.e., $v_k=v$), the uncertainty of $v$ converges to the total SNR:
\begin{align}
    \lim_{v_k\xrightarrow{}0}\sigma^{-2}(v) = \sum_k \lim_{v_k\xrightarrow{}0}\sigma^{-2}(v_k) = 2\sum_k\sum_{\ell}^{\ell_{\rm max}}(2\ell+1)(1+\zeta^{-1}_{\ell,k})^{-2} = \SNR^2_{\rm (ast)}.\label{eq:sigmavEB}
\end{align}
Note that this assumption applies to cosmological sources rather than a finite sum of nearby SMBHBs. 

In the top panel of Fig.\,\ref{fig:Posterior_Astro}, we display the posterior distribution on the $\{\log_{10}A,v\}$ plane for Gaia and {XG-Gaia}, assuming a truth value of $v=0.1$. The degeneracy between the two parameters turns out to be negligible. We also present the marginalized distribution of both parameters, finding that the uncertainty for {XG-Gaia}, $\sigma(v)\simeq 0.022$, is very close to $1/\SNR$. Thus, {XG-Gaia} serves as a powerful test for circular polarization that can reach a fractional value as low as $2.2\%$. In contrast, the marginal detection of SGWB for Gaia ($\text{SNR}\simeq 1.5$) makes it challenging to resolve the circular polarization component.

\subsubsection{Synergistic Analyses between PTAs and Astrometry}

In this section, we explore the synergistic potential of joint observations involving both PTAs and astrometry, considering all possible correlations. Assuming an ideal scenario where all measurements share the same cadence and total observation time, we consider independent measurements $\hbv{X}_{\ell m,k}=\{z_{\ell m,k},\,E_{\ell m,k},\,B_{\ell m,k}\}$ with a covariance matrix $\hbv{\Sigma}_{\ell,k}$ defined in Eq.\,(\ref{eq:CN}). Within $\hbv{\Sigma}_{\ell,k}$, $C_{(h)\ell,k}^{zz/EE/BB/zE}$ are sensitive to $I_k$, while $C_{(h)\ell,k}^{zB/EB}$ are proportional to $V_k$. Inserting $\hbv{\Sigma}_{\ell,k}$ into the likelihood Eq.\,(\ref{Eq:lnlike_general}), the SNR in the limit where $v_k\equiv V_k/I_k=0$ is given by:
\begin{align}
 &\SNR^{2}_{{\rm (syn)}} =\sum_k \sum_{\ell=2}^{\ell_{\rm max}^{\rm (syn)}}(2\ell+1)\frac{\zeta ^2 \left(2 \beta ^2 (\zeta +1)^2+2 \beta  (2 \zeta +1) (\zeta +1)+2 \zeta  (\zeta +1)+1\right)}{(\zeta +1)^2  (\beta  \zeta +\beta +\zeta )^2}\label{Eq:Conjoint_SNR},
\end{align}
where $\beta_{\ell,k}\equiv  B_{\ell}^2{(S^{(n)}_{\delta z,k}/N_{\delta z})}/{(S^{(n)}_{\delta \hbv{x},k}/N_{\delta \hbv{x}}})$, and we omit the index $\ell$ and $k$ on the right-hand side for simplicity. We keep 
$\ell_{\rm max}^{\rm (syn)}$ to be the one of PTA, as $\ell_{\rm max}$ for astrometry is typically much higher.

In the weak signal region, where $\zeta_{\ell,k} \equiv I_kA_{\ell}B_{\ell}^2/(4\pi S_{\delta \hbv{x},k}^{(n)}/N_{\delta \hbv{x}}) \ll 1$, the SNR simplifies to:
\begin{align}
         &\SNR_{\rm (syn)}^{2} \xrightarrow{\zeta_{\ell,k}\ll1} \sum_{k}\sum_{\ell=2}^{\ell_{\rm max}^{\rm (syn)}} (2\ell+1)\,(
         1/\beta_{\ell,k}^2+2/\beta_{\ell,k}+2
         )\,\zeta_{\ell,k}^2.
\end{align}
Here, the $\beta^{-2}$, $\beta^{-1}$, and $\beta^0$ terms correspond to the estimation from PTA, PTA-astrometry correlation, and astrometry only, respectively. Thus, in addition to the sum of each type of observation, the synergistic observation has the additional contribution from the cross-correlation among the two.

As astrometry has a higher $\ell_{\rm max}$ than PTAs, the remaining astrometric-only $\ell$-modes can be included separately into the analysis:
\be
\begin{aligned}
&\ln\mathcal{L}_{\rm (tot)} = \ln\mathcal{L}_{\rm (syn)}+ \left.\ln\mathcal{L}_{\rm (ast)}\right|_{\ell_{\rm max}^{\rm (syn)}<\ell\leq \ell_{\rm max}^{\rm (ast)}},\\
    &\SNR_{\rm (tot)}^{2} = \SNR_{\rm (syn)}^{2} + \left.\SNR_{\rm (ast)}^{2}\right|_{\ell_{\rm max}^{\rm (syn)}<\ell\leq\ell_{\rm max}^{\rm (ast)}}.\label{Eq:Tot_SNR}
\end{aligned}\ee
Here, we define $\ell_{\rm max}^{\rm (syn)}$ and $\ell_{\rm max}^{\rm (ast)}$ to distinguish the maximal $\ell$ from the synergistic analysis and that from the astrometry-only.

\paragraph{Parameter Estimation in Synergistic Analyses of PTA and Astrometry}
We explore two pairs of joint observations: the ongoing NANOGrav\,+\,Gaia and the future SKA\,+\,{XG-Gaia}. A challenge in these synergistic analyses arises from differences in cadence and total observation time, as detailed in Eq.\,(\ref{Eq:PTA_obs}) and (\ref{Eq:Gaia_obs}). In our approach, we adopt a conservative strategy by selecting the longer cadence, the shorter observation time, and the smaller value among $N_z$ and $N_{\delta \hbv{x}}$ from each pair. The corresponding benchmark parameters are as follows:
\be
\begin{aligned}
\text{NANOGrav\,+\,Gaia}:&\qquad \Delta t = 24\, {\rm days},\ T_{\rm obs} = 10\,{\rm yr},\ N_{z} = N_{\delta \hbv{x}} = 50,\\
\text{SKA\,+\,{XG-Gaia}}:&\qquad \Delta t = 24\, {\rm days},\ T_{\rm obs} = 20\,{\rm yr},\ N_{z} = N_{\delta \hbv{x}} = 200.
\label{Eq:Con_definition1}
\end{aligned}\ee
The measurement uncertainties and the total number of observed stars remain consistent with Eq.\,(\ref{Eq:PTA_obs}) and (\ref{Eq:Gaia_obs}).

The total SNR and parameter resolution for the power-law SGWB model are comparable between the more sensitive of the pair, namely NANOGrav and {XG-Gaia}. Thus, no additional figures similar to Fig.\,\ref{fig:SNR_PTA} and \ref{fig:sensitivity_para} are presented for further comparison.

The measurement of circular polarization in PTA-astrometry cross-observations can be realized through the $zB$ correlation, in addition to the $EB$ correlation in the astrometry-only correlations. Similar to the process outlined in Sec.\,\ref{sec:ast}, we first calculate the  information matrix $\mathcal{I}_{k}^{ij}$ and derive the corresponding uncertainty of $v$. Due to the complexity of the expressions, we present the result for the case when $|v|\ll1$:
\begin{align}
     \lim_{v\rightarrow 0}\sigma^{-2}(v)  = 2\sum_k\sum_{\ell=2}^{\ell_{\rm max}}(2\ell+1) \frac{ (\beta_{\ell,k} +1) \zeta_{\ell,k} ^2}{(\zeta_{\ell,k} +1) (\beta_{\ell,k}  \zeta_{\ell,k} +\beta_{\ell,k} +\zeta_{\ell,k} )}.\label{Eq:Tot_Circ}
\end{align}
This expression is the linear sum of contributions from both the $zB$ correlation:
\begin{align}
   \lim_{v\rightarrow 0} \sigma^{-2}_{zB}(v)  = 2\sum_k\sum_{\ell=2}^{\ell_{\rm max}}(2\ell+1) \frac{\zeta_{\ell,k}^2}{(\zeta_{\ell,k} +1)^2(\beta_{\ell,k}  \zeta_{\ell,k} +\beta_{\ell,k} +\zeta_{\ell,k} )},\label{Eq:Circ_zB}
\end{align}
and the $EB$ correlation in Eq.\,(\ref{eq:sigmavEB}).

In the bottom panel of Fig.\,\ref{fig:Posterior_Astro}, we present the posterior distribution for $\{\log_{10}A, v\}$ for the two joint observations, along with the two astrometry-only observations. The SKA\,+\,{XG-Gaia} pair exhibits slightly better resolution $\sigma(v)\simeq0.022$ compared to the {XG-Gaia}-only observation, attributed to {XG-Gaia}'s overall better sensitivity. On the other hand, the NANOGrav\,+\,Gaia pair can achieve a much superior resolution with $\sigma(v)\simeq0.39$ compared to the Gaia-only observation, benefiting from NANOGrav's higher sensitivity. Thus, we conclude that to effectively constrain circular polarization using current-generation observations, a cross-correlation between PTA and Gaia is necessary.

\clearpage

%%%%%%%%%%%%%%%%%%%%%%%%%%%%%%%%%%%%%%%%%%%%%%%%%
\section{Emergence of Generalized Hellings-Downs Correlation Patterns}\label{sec:correlation}
%%%%%%%%%%%%%%%%%%%%%%%%%%%%%%%%%%%%%%%%%%%%%%%%%

An essential feature for identifying the quadrupolar nature of SGWB lies in the spatial correlation pattern, which can be revealed through either the coefficients of $C_{(h)\ell}^{XX'}$ as a function of $\ell$, as discussed in Sec.\,\ref{sec:SH}, or in the separation angle space, such as the Hellings-Downs curve recently explored by various PTA collaborations~\cite{NANOGrav:2023gor,EPTA:2023fyk,Reardon:2023gzh,Xu:2023wog}. While we extensively discussed observables in spherical harmonic space in Sec.\,\ref{sec:SNR}, this section focuses on correlation functions in the separation angle space.

Instead of using the information matrix, we generate random realizations of both SGWB-induced redshift/deflections and measurement noises. These realizations ultimately lead to predictions of spatial correlation patterns with uncertainties in each separation angle bin.

%%%%%%%%%%%%%%%%%%%%%%%%%%%%%%%%%%%%%%%%%%%%%%%%%
\subsection{Realization of Correlations in PTA and Astrometry}
%%%%%%%%%%%%%%%%%%%%%%%%%%%%%%%%%%%%%%%%%%%%%%%%%

In this section, we elaborate on the detailed methodology employed for simulating SGWB-induced signals in both PTA and astrometric observations in configuration space, presenting various illustrative examples of results.

To commence, we partition the celestial sphere into $N$ evenly distributed patches using \texttt{HEALPix}~\cite{Gorski:2004by}, denoting their central locations as $\{\hbf{n}_a\}$. In the frequency domain, the signals are complex variables. Each patch is attributed a stochastic complex dimension-$3$ vector denoted as $(\delta z_a,\delta \hbv{x}_a)$, where $\delta \hbv{x}_a$ represents complex dimension-$2$ vectors on planes perpendicular to each $\hbf{n}_a$. The generation of $\delta z_a$ and $\delta \hbv{x}_a$ follows a probability distribution given by
\begin{align}
   \left( \{\delta z_a\}, \{\delta\hbv{x}_a\}\right)\sim \mathcal{NC}(\hbv{0}_{3N},{\hbv{C}}),\qquad \hbv{C} = \hbv{C}_{(h)} + \hbv{C}_{(n)}.
\end{align}
Here, $\hbv{C}$ is the $3N\times3N$ complex covariance matrix, representing the linear sum of the SGWB-induced correlation $\hbv{C}_{(h)}$ and the noise $\hbv{C}_{(n)}$. The SGWB covariance matrix, following Ref.~\cite{Mihaylov:2018uqm}, is defined as:
\begin{align}
    \hbv{C}_{(h)}(I,V)  \equiv \begin{pmatrix}
        \hbv{C}_{\delta z}&\hbv{C}_{\delta z\delta \hbv{x}}\\
        \hbv{C}_{\delta z\delta \hbv{x}}^{\dagger}&\hbv{C}_{\delta \hbv{x}}\\
    \end{pmatrix},
\end{align}
where $I$ and $V$ are the true parameters of total intensity and circular polarization of the SGWB at a given frequency (with the frequency label $k$ omitted for simplicity). The matrices $\hbv{C}_{\delta z}$, $\hbv{C}_{\delta z\delta \hbv{x}}$, and $\hbv{C}_{\delta \hbv{x}}$ are of dimensions $N\times N$, $N\times 2N$, and $2N\times 2N$, respectively. Their definitions mirror those of Eq.\,(\ref{eq:HD}) and (\ref{eq:ORFA}) but lack the $\delta$-functions:
\be
\begin{aligned}
    \hbv{C}_{\delta z}^{ab} \equiv&\, \Gamma_{z}(\theta_{ab})\,I,\\
        \hbv{C}^{ab}_{\delta z\delta \hbv{x}} \equiv&\, \Gamma_{z\hbv{\delta x}}(\theta_{ab})\left[
    I\,\hbf{e}_{||}^b
    +iV\,\hbf{e}_{\perp}
    \right],\\
    \hbv{C}^{ab}_{\delta\hbv{x}} \equiv&\, \Gamma_{\delta \hbv{x}}(\theta_{ab})\left[I \left(\hbf{e}_{||}^a\hbf{e}_{||}^b+\hbf{e}_{\perp}\hbf{e}_{\perp}\right) + iV \left(\hbf{e}_{||}^a\hbf{e}_{\perp}-\hbf{e}_{\perp}\hbf{e}_{||}^b\right)\right].
\end{aligned}
\ee
Here, $\hbf{e}_{||}^a$, $\hbf{e}_{||}^b$, and $\hbf{e}_{\perp}$ are defined in Eq.\,(\ref{Eq:unit_vectors}), and the $\Gamma$ functions are defined in Eqs.\,(\ref{eq:HD}) and (\ref{Eq:correlations}). The symmetry $(\hbv{C}_{\delta \hbv{x}}^{ab})^{\dagger} = \hbv{C}_{\delta \hbv{x}}^{ba}$ ensures that $\hbv{C}_{\delta x}$ is Hermitian, and so is the covariance matrix $\hbv{C}_{(h)}$.

The noise matrix is purely diagonal due to spatially uncorrelated measurement noise:
\begin{align}
    {\hbv{C}}_{(n)}={\rm diag} \left( S^{(n)}_z\boldsymbol{1}_{N}, S^{(n)}_{\delta \hbv{x}}\boldsymbol{1}_{2N}\right),
\end{align}
where $S^{(n)}_{z}$ and $S^{(n)}_{\delta \hbv{x}}$ are defined in Eq.\,(\ref{Eq:PTA_RN}) and (\ref{eq:Sndeltax}), respectively.

In practice, to generate data from a complex normal distribution, we decompose both the signals and the covariance matrix into real and imaginary parts:
\be   \left( \{{\Re}[ \delta z_a]
\}, \{ {\Re}[\delta\hbv{x}_a]\}, \{{\Im}[ \delta z_a]
\}, \{ {\Im}[\delta\hbv{x}_a]\}\right) \sim \mathcal{N}(\boldsymbol{0}_{6N},\hbv{{C}}'),\quad \hbv{{C}}' \equiv \frac{1}{2}\begin{pmatrix}
        {\Re}[\hbv{{C}}]&{\Im}[\hbv{{C}}]^T\\
       {\Im}[\hbv{{C}}] &{\Re}[\hbv{{C}}]
    \end{pmatrix}. \ee
Here, ${\Re}[\hbv{{C}}]$ contains signals proportional to $I$ and measurement noise ${\hbv{C}}_{(n)}$, while ${\Im}[\hbv{{C}}]$ only contains the term proportional to $V$.

\begin{figure}[h]
    \centering
    \includegraphics[width=1.\textwidth]{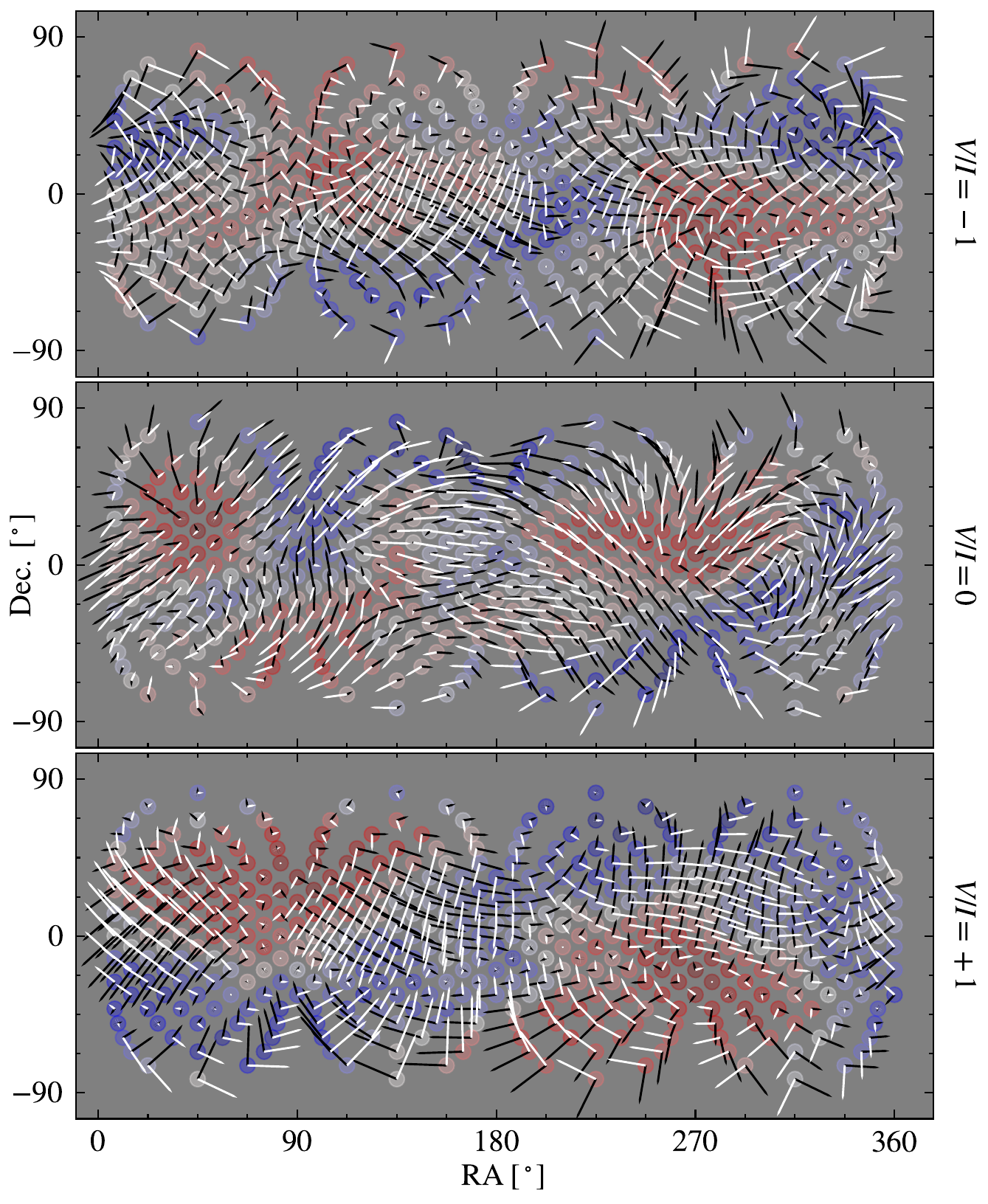}
    \caption{Examples of realizations on the celestial sphere for $V/I = -1$ (top), $0$ (middle), and $1$ (bottom), featuring a total of $432$ patches.  Each patch contains information on the real part of the redshift $\delta z$ represented on a circle, spanning from red ($>0$) to blue ($<0$). The real (black arrow) and imaginary (white arrow) components, with lengths proportional to their magnitudes, are also displayed. The measurement noise is not included in these examples. In cases with $V/I = \pm 1$, the real and imaginary arrows are always perpendicular, while in the unpolarized case, they are uncorrelated.}  
    \label{fig:realizations}
\end{figure}

In Fig.\,\ref{fig:realizations}, we present three cases of realizations for both redshift $\delta z$ and angular deflection $\delta\hbv{x}$ with $V/I = -1, 0,$ and $1$, respectively. The background circle colors represent the real part of $\delta z$, ranging from red $(\delta z>0)$ to blue $(\delta z<0)$. The real and imaginary components of the angular deflections are depicted with black and white arrows, respectively, with lengths proportional to their magnitudes. In these examples, noise is assumed to vanish. Consequently, in the two maximally polarized cases, the real and imaginary parts of $\delta \hbv{x}$ are always perpendicular, with the relative phases having different signs for the two cases. On the other hand, for $V=0$, they exhibit random behavior without any correlations.

\begin{figure}[h]
    \centering
    \includegraphics[width=\textwidth]{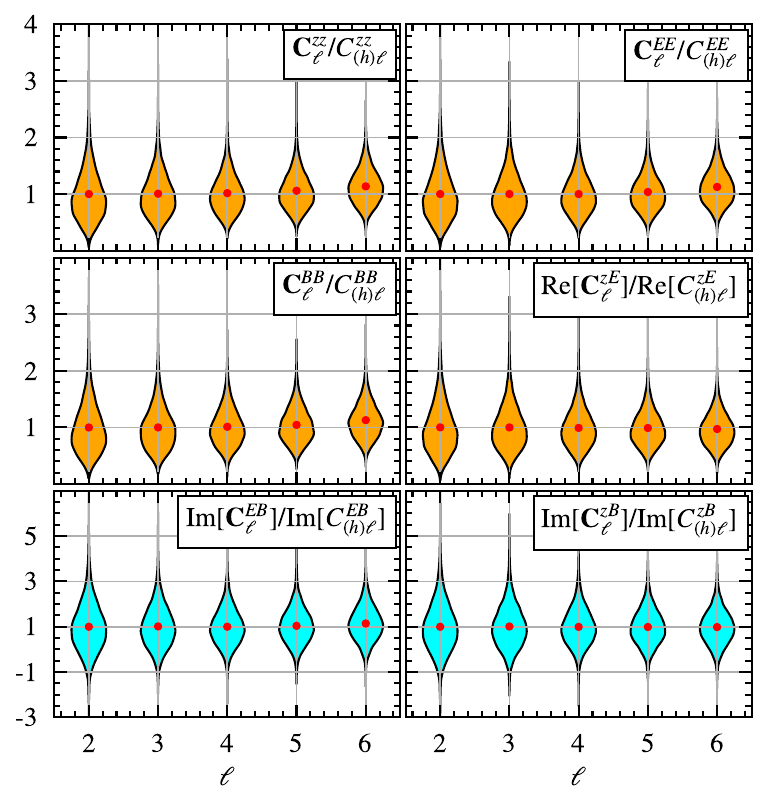}
    \caption{Reconstructions of rotationally invariant power spectra in the spherical harmonic space for $\ell$ ranging from $2$ to $6$. Each reconstruction is derived from a random realization of a map of $108$ patches of $\{\delta z_a\}$ and $\{\delta\hbv{x}_a\}$ using Eqs.\,(\ref{eq:zEBr}) and (\ref{Eq:def_cxx}), in the absence of measurement noise. Total intensity $I$ estimators are depicted in orange, while circular polarization $V$ estimators are shown in blue, assuming $V/I=0.3$. Each spectrum is normalized by its theoretical average value. Violins represent the variance of reconstructions from $10^4$ realizations, and the red dots denote the average values of realizations. Both are consistent with the variance calculation in the denominator of Eq.\,(\ref{eq:snr}).   
    }
    \label{fig:enter-label}
\end{figure}

Another consistency check, aiming to connect with the previous discussion in the spherical harmonic space in Sec.\,\ref{sec:SNR}, involves the reconstruction of the spherical harmonic observable from $\{z_a\}$ and $\{\delta\hbv{x}_a\}$ using Eqs.\,(\ref{eq:zEBr}) and (\ref{Eq:def_cxx}), in the absence of measurement noise. 
Examples of the corresponding $\hbv{C}_{\ell}^{XX'}$ are depicted in Fig.\,\ref{fig:enter-label}. The violins at each $\ell$-mode showcase the statistical distribution of the estimated $\hbv{C}_{\ell}^{XX'}$ values for all $\ell\leq 6$, based on a total of $10^4$ realizations. The modes for $\ell=0$ and $1$ turn out to be negligible, as expected. We normalize the remaining modes by their expected average values, involving $C^{XX'}_{(h)\ell}$ defined in Eqs.\,(\ref{Eq:ChXX_I}) and (\ref{Eq:ChXX_V}). Consequently, the red dots representing the average values consistently hover around $1$. The variance for $I$ estimators, including $zz, EE, BB,$ and $zE$ correlations, shown in orange, all exhibit uncertainties of approximately $1$. This is again consistent with the theoretical predictions in the denominator of Eq.\,(\ref{eq:snr}). For $V$ estimators involving $EB$ and $zB$ correlations, their variance, shown in blue, is larger than $1$, attributed to our assumption of $V/I=0.3$, and the fact that the variance for $V$ includes a contribution from $I$.

\clearpage

%%%%%%%%%%%%%%%%%%%%%%%%%%%%%%%%%%%%%%%%%%%%%%%%%
\subsection{Generalized Hellings-Downs Correlations}
%%%%%%%%%%%%%%%%%%%%%%%%%%%%%%%%%%%%%%%%%%%%%%%%%
In this section, we employ the developed simulation to project how spatial correlations in the configuration space manifest for various PTAs and astrometric observations, accounting for a more realistic scenario where both measurement noise and intrinsic variance of SGWB are present.

We consider distinct cases of observations: NANOGrav and SKA as PTAs, Gaia and {its next-generation upgrade} as astrometry missions, and their cross-correlations. Each observation channel differs in terms of the number of pulsars or patches, and noise levels, with corresponding benchmark parameters listed in Eqs.\,(\ref{Eq:PTA_obs}), (\ref{Eq:Gaia_obs}), and (\ref{Eq:Con_definition1}). For the simulations, we use the fiducial SGWB model defined in Eq.\,(\ref{Eq:Phi_to_PSD1}), assuming $V/I=0$ throughout the analysis.

We conduct a total of $100$ simulations, each comprising realizations for all frequency bins, resulting in distributions of $\{ \delta z_a, \delta \hbv{x}_a \}$ maps for each frequency. For every pair of patches, characterized by their separation angle $\theta$, we categorize them into $11$ evenly distributed bins spanning from $0$ to $180^\circ$. Within each separation angle bin, we compute the averages of the products $\delta z_a \delta z_b^*$, $(\delta{x}_a^{||} \delta{x}_b^{||*}+\delta{x}_a^{\perp} \delta{x}_b^{\perp*})/2$, and $\delta z_a \delta{x}_b^{||*}$, where $\delta{x}_a^{||/\perp}$ represents the projection of $\delta\hbv{x}_a$ along $\hbf{e}_{||}^a$ or $\hbf{e}_{\perp}$. This computation yields a sky-averaged two-point function for a single realization. From the $100$ simulations, we determine the standard deviation at the $\theta_i$-th separation angle and frequency mode $k$ as $\sigma_{\theta_i,k}$, incorporating contributions from both the SGWB and measurement noises.

\begin{figure}[h]
    \centering
    \includegraphics[width=1\textwidth]{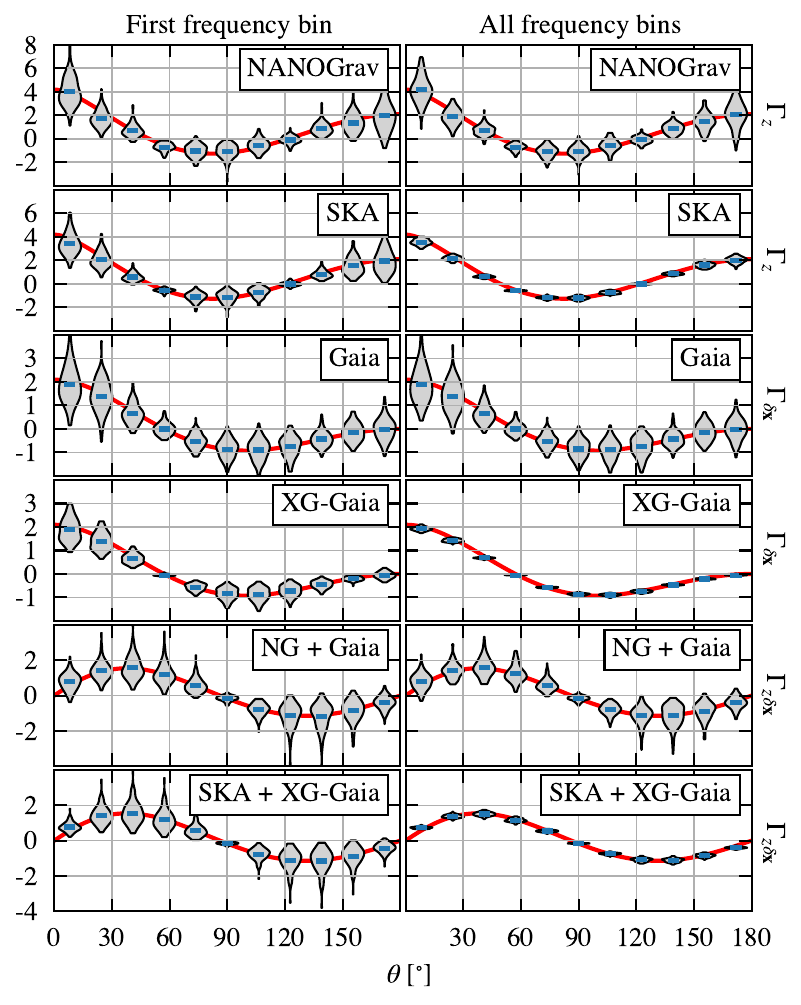}
        \caption{Distribution of sky-averaged two-point functions from $100$ simulations presented in gray violins for various PTA and astrometric observation channels, displayed for the first frequency bin (left) and weight-averaged across all frequency bins (right). The red lines illustrate the generalized Hellings-Downs curves defined in Eqs.\,(\ref{eq:HD}) and (\ref{Eq:correlations}). The variance from the simulations encompasses both SGWB variance and measurement noises. The central values of the violins, shown in blue, align with the red lines.}
    \label{fig:orfs}
\end{figure}

In the left part of Fig.\,\ref{fig:orfs}, violins are employed to illustrate the distribution from $100$ simulations of sky-averaged two-point functions for various observation channels, presenting only the first frequency bin of the observations. 
 The red solid lines represent the theoretical prediction for the average correlations, as defined in Eqs.\,(\ref{eq:HD}) and (\ref{Eq:correlations}).  In the right part, we conduct an average across all frequency bins for each separation angle, incorporating a weight factor $1/\sigma^2_{\theta_i,k}$. The frequency-averaged variance $\sigma_{\theta_i}^2$, defined as $1/\sigma^2_{\theta_i}\equiv \Sigma_k 1/\sigma^2_{\theta_i,k}$, becomes narrower, and the average values, shown in blue, closely align with the theoretical predictions.

\begin{figure}
    \centering
    \includegraphics[width=1\textwidth]{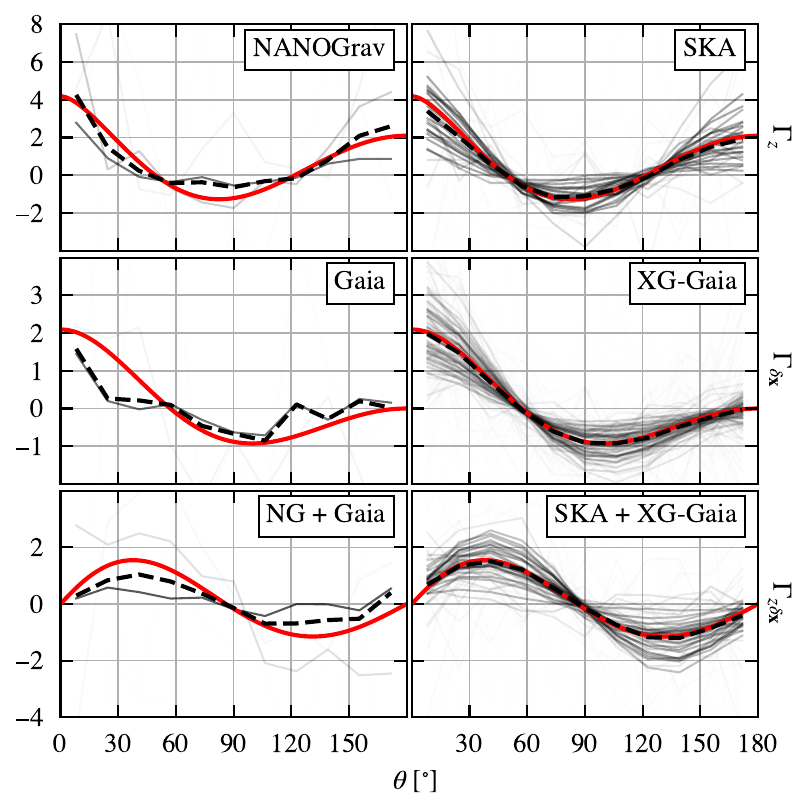}
    \caption{Sky-averaged two-point functions from a single simulation for various PTA and astrometric observation channels. Each gray line represents a realization in a specific frequency bin, with opacity inversely proportional to the variance associated with that frequency bin. The black dashed lines depict the weighted averages of all frequency bins, aligning with the generalized Hellings-Downs curves (red) defined in Eqs.\,(\ref{eq:HD}) and (\ref{Eq:correlations}) for observations involving SKA or {XG-Gaia}.}
    \label{fig:orfslines}
\end{figure}

Figure\,\ref{fig:orfslines} illustrates predictions for spatial correlations across various observation channels. Each gray line represents one realization obtained at a specific frequency bin. The opacity of each line corresponds to the weight factor of the frequency, defined as $1/\sigma^2_{k}\equiv \Sigma_i 1/\sigma^2_{\theta_i,k}$. The black dashed lines represent the averages of these gray lines, accounting for the weight factor $1/\sigma^2_{\theta_i,k}$ across all frequencies. For current observations involving NANOGrav or Gaia, the gray lines fluctuate around the generalized Hellings-Downs curves in red, while the dashed lines exhibit some deviation from it. In contrast, next-generation observations like SKA or {XG-Gaia} have averaged values that align well with the red. The gray lines are predominantly localized in specific regions, a result of the stochastic nature of the SGWB rather than measurement noise, owing to the high sensitivity of these measurements. In the next subsection, we will delve into a detailed discussion of the origin of uncertainty regions, commonly referred to as cosmic variances.

\clearpage

%%%%%%%%%%%%%%%%%%%%%%%%%%%%%%%%%%%%%%%%%%%%%%%%%
\subsection{Cosmic Variances}
%%%%%%%%%%%%%%%%%%%%%%%%%%%%%%%%%%%%%%%%%%%%%%%%%

As depicted in Fig.\,\ref{fig:orfslines}, even in the limit of SNR\,$\gg1$ for high-sensitivity observations that encompass a large number of patches, the sky-averaged spatial correlations still exhibit an uncertainty envelope. The discussion of these inherent uncertainties for PTAs, referred to as cosmic variance, has been the subject of recent studies in Refs.~\cite{Allen:2022dzg,Allen:2022ksj,Bernardo:2022xzl,Bernardo:2023bqx,Bernardo:2023pwt,Romano:2023zhb,Bernardo:2023zna}. In this section, we present a theoretical derivation of cosmic variances for general PTA and astrometric observations.

Cosmic variance manifests itself in patch pairs separated by a specific angular separation on the celestial sphere, denoted as the variance intrinsic to the sky-averaged spatial correlations at a given separation angle $\theta$. To calculate cosmic variance, we begin by introducing the sky-averaged two-point functions~\cite{Ng:1997ez,Bernardo:2022xzl}:
\be  \left\{ X\, X'^*\right\}^S_{\theta} \equiv \int \frac{d^2\hbf{n}_a}{4\pi} \int \frac{d^2\hbf{n}_b}{4\pi}\, X_a X'^*_b\, \delta\left(\hbf{n}_a\cdot\hbf{n}_b - \cos \theta\right), \label{eq:SAXX} \ee
where $X$ can represent $\delta z$, $\delta x^{||}$, or $\delta x^{\perp}$. This expression can be further simplified in the spherical harmonic space, as seen in the case of PTAs~\cite{Gair:2014rwa,Roebber:2016jzl,Roebber:2019gha,Qin:2018yhy,Bernardo:2022xzl,Nay:2023pwu,Liang:2023ary}:
\begin{equation}
    \begin{aligned}
        \left\{ \delta z\, \delta z^*\right\}^S_{\theta}  &= \sum_{\ell,m}\sum_{\ell'm'} z_{\ell m}z_{\ell'm'}^* \int \frac{d^2\hbf{n}_a}{4\pi} \int \frac{d^2\hbf{n}_b}{4\pi}\, Y_{\ell m}(\hbf{n}_a) Y^*_{\ell'm'}(\hbf{n}_b) \,\delta\left(\hbf{n}_a\cdot\hbf{n}_b - \cos \theta\right) \\
        &= \sum_{\ell}\frac{2\ell + 1}{4\pi}\,\hbv{C}_{\ell}^{zz}\,P_\ell(\cos \theta).
        \label{eq:zzSA}
    \end{aligned}
\end{equation}
Here, $P_\ell$ are the Legendre polynomials.

A parallel set of steps can be applied to astrometric observations, taking into account their expansion in terms of vector spherical harmonics as follows~\cite{Qin:2018yhy}:
\begin{equation}
    \begin{aligned}
        &\left\{ \delta x^{||}\, \delta x^{||*}\right\}^S_{\theta}\\ =& \sum_{X,X'\in \{E, B\}}\sum_{\ell,m}\sum_{\ell'm'} X_{\ell m}X_{\ell'm'}^{'*} \int \frac{d^2\hbf{n}_a}{4\pi} \int \frac{d^2\hbf{n}_b}{4\pi}\, 
        \left[{\bf Y}_{\ell m}^X(\hbf{n}_a)\right]_{||} \left[{\bf Y}_{\ell'm'}^{X'}(\hbf{n}_b)\right]^*_{||}\,\delta\left(\hbf{n}_a\cdot\hbf{n}_b - \cos \theta\right) \\
        =& \sum_{\ell}\frac{2\ell + 1}{4\pi} \left[{\bf C}_\ell^{EE} G_{(\ell 1)}(\theta) + {\bf C}_\ell^{BB} G_{(\ell 2)}(\theta)\right].\label{eq:ppSA}
    \end{aligned}
\end{equation}
Here,
\begin{equation}
    \begin{aligned}
        G_{(\ell 1)}(\theta) \equiv -\frac{1}{2}\left[\frac{1}{\ell(\ell + 1)}P_\ell^2(\cos \theta) - P_\ell^0(\cos \theta)\right], \qquad
        G_{(\ell 2)}(\theta) \equiv -\frac{1}{\ell(\ell + 1)}\frac{P_\ell^1(\cos \theta)}{\sin \theta},
    \end{aligned}
\end{equation}
are functions of associated Legendre polynomials $P_\ell^m$~\cite{Qin:2018yhy}. The expression for $\left\{ \delta x^{\perp}\, \delta x^{\perp*} \right\}^S_{\theta}$ only differs from Eq.\,(\ref{eq:ppSA}) by switching $G_{(\ell 1)}\leftrightarrow G_{(\ell 2)}$. The expression for the PTA-astrometry cross-correlation can be derived similarly:
\be \left\{ {\Re}\left[ \delta z\, \delta{x}^{||*} \right] \right\}^S_{\theta} =  \sum_{\ell}\frac{2\ell + 1}{4\pi}\,\frac{1}{\sqrt{\ell(\ell+1)}}\, {\Re}\left[\hbv{C}_{\ell}^{zE}\right]\,P^1_\ell(\cos \theta). 
\label{eq:zPSA}\ee
The explicit forms of ${\bf C}_\ell^{zz}, {\bf C}_\ell^{EE}$, ${\bf C}_\ell^{BB}$, and  ${\bf C}_\ell^{zE}$ have been defined in Eq.\,(\ref{Eq:def_cxx}). The ensemble averages of Eqs.\,(\ref{eq:zzSA}), (\ref{eq:ppSA}), and (\ref{eq:zPSA}) yield the generalized Hellings-Downs curves as defined in Eqs.\,(\ref{eq:HD}) and (\ref{Eq:correlations}).

The cosmic variance (CV) at a separation angle precisely corresponds to the variance of the sky-averaged two-point correlations defined in Eq.\,(\ref{eq:SAXX}):
\begin{equation}\label{eq: CV_definition}
    {\rm CV}(X\,X'^*)_{\theta} = \left\langle \left(\left\{ X X'^*\right\}^S_\theta\right)^2 \right\rangle - \left\langle \left\{ X X'^*\right\}^S_{\theta} \right\rangle^2.
\end{equation}
Here, $\langle \cdots \rangle$ denotes the ensemble average over the SGWB, as the definition of CV does not include measurement noises. The calculation of CV can be simplified using correlations in the spherical harmonic space~\cite{Bernardo:2022xzl}.

For the PTA-only observation, the CV becomes:
\begin{equation}
\begin{aligned}
    {\rm CV}({\delta z\, \delta z^*})_{\theta} =& \sum_{\ell\ell'}\frac{(2\ell + 1)}{4\pi}\frac{(2\ell' + 1)}{4\pi} P_\ell(\cos \theta)P_{\ell'}(\cos \theta)\vev{{\bf C}_\ell^{zz}{\bf C}_{\ell'}^{zz}} - \left(\sum_{\ell}\frac{2\ell + 1}{4\pi}P_\ell(\cos \theta)\vev{{\bf C}_\ell^{zz}}\right)^2\\
=& \sum_{\ell} \frac{2\ell + 1}{16\pi^2} \left(C_{(h)\ell}^{zz}\, P_\ell(\cos \theta)\right)^2.\label{eq:CVzz}
    \end{aligned}
\end{equation}
In this calculation, we used $\vev{{\bf C}_\ell^{XX'}} = C_{(h)\ell}^{XX'}$ and 
\begin{equation}
    \vev{{\bf C}_\ell^{XX}{\bf C}_{\ell'}^{XX}} = C_{(h)\ell}^{XX}C_{(h)\ell'}^{XX} + \frac{1}{2\ell + 1}\left(C_{(h)\ell}^{XX}\right)^2\delta_{\ell\ell'},\label{eq:ITX}
\end{equation}
which arises from Isserlis' theorem~\cite{10.1093/biomet/11.3.185}.

A similar result can be obtained for the astrometry-only correlations. We begin by calculating the first term in Eq.\,(\ref{eq: CV_definition}), focusing on the parallel directions:
\begin{equation}
    \begin{split}
     \left\langle {\left(\left\{ \delta x^{||} \delta x^{||*} \right\}^S_{\theta}\right)^2} \right\rangle = \sum_{\ell\ell'}\frac{(2\ell + 1)(2\ell' + 1)}{16\pi^2} [&\vev{{\bf C}_\ell^{EE}{\bf C}_{\ell'}^{EE}}G_{(\ell 1)}G_{(\ell' 1)} + \vev{{\bf C}_\ell^{BB}{\bf C}_{\ell'}^{BB}}G_{(\ell 2)}G_{(\ell' 2)} + \\ &\vev{{\bf C}_\ell^{EE}{\bf C}_{\ell'}^{BB}}G_{(\ell 1)}G_{(\ell' 2)} + \vev{{\bf C}_\ell^{BB}{\bf C}_{\ell'}^{EE}}G_{(\ell 2)}G_{(\ell' 1)}]\,,
    \end{split}
\end{equation}
where we omit the $\theta$-dependence in  $G_{(\ell1)}(\theta)$ and $G_{(\ell 2)}(\theta)$ for simplicity. The first two terms within the $[\cdots]$ can be computed directly from Eq.\,(\ref{eq:ITX}), while the last two terms represent
\begin{equation}
        \vev{{\bf C}_\ell^{EE}\, {\bf C}_{\ell'}^{BB}} = C_{(h)\ell}^{EE} \,C_{(h)\ell'}^{BB},\qquad
        \vev{{\bf C}_\ell^{BB}\, {\bf C}_{\ell'}^{EE}} = C_{(h)\ell}^{BB}\,C_{(h)\ell'}^{EE},
\end{equation}
in the absence of circular polarization. The second term in Eq.\,(\ref{eq: CV_definition}) is merely the square of the sky average in Eq.\,(\ref{eq:ppSA}) after applying $\vev{{\bf C}_\ell^{XX'}} = C_{(h)\ell}^{XX'}$. Combining these two terms yields the final CV result:
\begin{equation}
    {\rm CV}(\delta x^{||}\, \delta x^{||*})_{\theta} = \sum_{\ell}\frac{2\ell + 1}{16\pi^2}
  \left(C_{(h)\ell}^{EE}\right)^2  
    \left( G_{(\ell 1)}(\theta)^2 + G_{(\ell 2)}(\theta)^2 \right),
\label{eq:CVPP}
\end{equation}
where we utilized the relationship $C_{(h)\ell}^{EE} = C_{(h)\ell}^{BB}$. The expressions for correlations involving $\delta x^\perp$ are derived analogously, with the result given by the above equation and $G_{(\ell 1)}\leftrightarrow G_{(\ell 2)}$ swapped.

Note that in Figs.\,\ref{fig:orfs} and \ref{fig:orfslines}, the astrometric observations involve a linear combination of parallel and perpendicular correlations, specifically $(\delta{x}_a^{||} \delta{x}_b^{||*}+\delta{x}_a^{\perp} \delta{x}_b^{\perp*})/2$. The CV for this configuration is 
\begin{equation}
    {\rm CV}\left( (\delta{x}^{||}\, \delta{x}^{||*}+\delta{x}^{\perp} \delta{x}^{\perp*})/2 \right)_{\theta} = \sum_{\ell}\frac{2\ell + 1}{16\pi^2}
  \left( C_{(h)\ell}^{EE} \right)^2
    \left(G_{(\ell 1)}(\theta) + G_{(\ell 2)}(\theta)\right)^2,
\label{eq:CVPandP}
\end{equation}
which differs from the parallel-only case presented in Eq.\,(\ref{eq:CVPP}).

Finally, turning our attention to the PTA-astrometry cross-correlation, its CV can be obtained through analogous procedures, yielding:
\begin{equation}
    {\rm CV}({ {\Re}\left[ \delta z\, \delta{x}^{||*} \right]})_{\theta} = \sum_{\ell} \frac{2\ell + 1}{16\pi^2} \frac{1}{\ell(\ell+1)} \left(C_{(h)\ell}^{zE}\, P^1_\ell(\cos \theta)\right)^2.\label{eq:CVzE}
\end{equation}
It is noteworthy that the CVs presented in Eqs.\,(\ref{eq:CVzz}), (\ref{eq:CVPandP}), and (\ref{eq:CVzE}) share a common characteristic: each $\ell$-mode is the square of the coefficients in their corresponding average values of Eqs.\,(\ref{eq:zzSA}), (\ref{eq:ppSA}), and (\ref{eq:zPSA}), divided by $(2\ell+1)$. The numerical values of these CVs align with the variance envelope observed in the right part of Fig.\,\ref{fig:orfslines}.

%%%%%%%%%%%%%%%%%%%%%%%%%%%%%%%%%%%%%%%%%%%%%%%%%
\section{Conclusion}\label{sec:disucssion}
%%%%%%%%%%%%%%%%%%%%%%%%%%%%%%%%%%%%%%%%%%%%%%%%%

In this study, we utilize a joint likelihood that incorporates both astrometric and PTA observations to predict the detection of SGWB and the resolution of SGWB parameters. Our analysis takes advantage of the diagonal structure inherent in both measurement noise and intrinsic SGWB variance in the spherical harmonic space. This results in an analytical framework that facilitates the comparison of various PTA and astrometric observations, providing intuitive insights. Astrometry showcases its advantages in the harmonic space: the abundance of stars allows for a high value of $l_{\rm max}$, yielding more independent estimators and, consequently, higher SNR in the strong signal regions. This SNR, in turn, translates into the resolution of SGWB parameters, enabling the precise dissection of SGWB properties.

Our findings yield several predictions regarding astrometric observations. Firstly, the upcoming full data release of Gaia is expected to marginally detect the SGWB, which has already been observed by current PTA observations. Consequently, Gaia can serve as a valuable cross-check for PTA results, providing an independent demonstration of spatial correlation distinct from PTA's Hellings-Downs curve. Gaia's individual observations can also contribute to checking the chirality nature of SGWB. However, a significantly improved resolution of chirality can be achieved through cross-correlations between current PTAs and Gaia. {The next-generation upgrade of Gaia} is poised to deliver the best-ever sensitivity to the SGWB, particularly for its spectrum and chirality. Precise measurements of these quantities are crucial for understanding the evolution of SMBHBs, including eccentricity distribution and environmental effects that may influence the low-frequency end of the spectrum, as well as any potential cosmological signatures. The exploration of the high-frequency turning point in the SGWB spectrum remains an intriguing area, where high-cadence observations by Roman can provide valuable insights.

Looking forward, the prospects for the field appear promising. The anisotropy in the SGWB, a facet beyond the scope of our current analysis, represents another crucial aspect that astrometric observations could potentially illuminate. Our assumption of uniformly distributed stars on the celestial sphere, with uniform noise levels and cadences, can be refined by considering the realistic star distribution found in Gaia datasets. Incorporating this distribution along with the response functions of astrometry and PTA further strengthens the complementarity between these two types of observations, as they exhibit sensitivity to different incoming directions of the GWs. Such angular-dependent sensitivity will play an essential role in resolving individual SMBHBs, likely marking the next significant milestone in gravitational astronomy.

\hspace{5mm}
%%%%%%%%%%%%%%%%%%%%%%%%%%%%%%%%%%

%%%%%%%%%%%%%%%%%%%%%%%%%%%%%%%%%%
\acknowledgments
%%%%%%%%%%%%%%%%%%%%%%%%%%%%%%%%%%
We are grateful to Neil Cornish and Marc Kamionkowski for their useful discussions, and to Yijun Wang for her valuable email exchanges.
X.X.~and I.S.~are supported by  Deutsche Forschungsgemeinschaft under Germany’s Excellence Strategy EXC2121 “Quantum Universe” - 390833306.
Y.C.~acknowledges support from the Villum Investigator program supported by the VILLUM Foundation (grant no.~VIL37766) and the DNRF Chair program (grant no.~DNRF162) by the Danish National Research Foundation. This project has received funding from the European Union's Horizon 2020 research and innovation programme under the Marie Sklodowska-Curie grant agreement No 101131233 and by~FCT (Fundação para a Ciência e Tecnologia I.P, Portugal) under project No.~2022.01324.PTDC.
X.X.~and Y.C.~acknowledge the support of the European Consortium for Astroparticle Theory in the form of an Exchange Travel Grant. X.X.~would like to express special thanks to the Mainz Institute for Theoretical Physics (MITP) of the Cluster of Excellence PRISMA*(Project ID 39083149) for its hospitality and support. L.D. acknowledges research grant support from the Alfred P. Sloan Foundation (Award Number FG-2021-16495), and support of Frank and Karen Dabby STEM Fund from the Society of Hellman Fellows.
M.\c{C}.~and N.A.K~acknowledge support from the John Templeton Foundation through Grant No.~62840.
M.\c{C}.~is also supported by NSF Grants No.~AST-2006538, PHY-2207502, PHY-090003 and PHY-20043, NASA Grants No.~20-LPS20-0011 and 21-ATP21-0010, and by Johns Hopkins University through the Rowland Research Fellowship.
This work was in part carried out at the Advanced Research Computing at Hopkins (ARCH) core facility (rockfish.jhu.edu), which is supported by the NSF Grant No.~OAC1920103.

%\bibliography{references.bib}
%\bibliographystyle{JHEP}

\providecommand{\href}[2]{#2}\begingroup\raggedright\endgroup

\end{document}